\newcommand{\be}{\begin{equation}}
\newcommand{\bea}{\begin{eqnarray}}
\newcommand{\ee}{\end{equation}}
\newcommand{\eea}{\end{eqnarray}}
\newcommand{\nn}{\nonumber}

\newcommand{\qa}{\alpha}
\newcommand{\qb}{\beta}

\newcommand{\qd}{\delta}
\newcommand{\qD}{\Delta}
\newcommand{\qe}{\varepsilon}

\newcommand{\qz}{\zeta}

\newcommand{\qk}{\kappa}
\newcommand{\ql}{\lambda}
\newcommand{\qL}{\Lambda}

\newcommand{\qr}{\rho}
\newcommand{\qs}{\sigma}

\newcommand{\qt}{\tau}
\newcommand{\qf}{\varphi}

\newcommand{\qJ}{\Psi}
\newcommand{\qo}{\omega}
\newcommand{\qO}{\Omega}

\newcommand{\sgn}{{\rm sgn}}

\newcommand{\tr}{{\rm tr}\,}

\newcommand{\tri}{\triangle}
\newcommand{\mod}{\;{\rm mod}\;}

\newcommand{\dagg}{^{\dag}}

\newcommand{\prt}{\partial}

\newcommand{\fr}[2]{{\textstyle \frac{#1}{#2}}}

\newcommand{\EE}{{\mathbb E}}

\newcommand{\RR}{{\mathbb R}}

\newcommand{\one}{{\mathbb 1}}
\newcommand{\sH}{{\sf H}}
\newcommand{\bits}{ \{0,1\} }
\newcommand{\Hmin}{{\sf H}_{\rm min}}

\newcommand{\cD}{{\mathcal D}}

\newcommand{\cG}{{\mathcal G}}

\newcommand{\cM}{{\mathcal M}}

\newcommand{\cO}{{\mathcal O}}

\newcommand{\cQ}{{\mathcal Q}}

\newcommand{\cT}{{\mathcal T}}
\newcommand{\cU}{{\mathcal U}}

\newcommand{\cX}{{\mathcal X}}
\newcommand{\cY}{{\mathcal Y}}

\newcommand{\pr}{{\rm Pr}}

\newcommand{\isdef}{\stackrel{\rm def}{=}}

\newcommand{\ket}[1]{| #1 \rangle}
\newcommand{\bra}[1]{\langle #1 |}
\newcommand{\inprod}[2]{\langle #1 | #2 \rangle}

\RequirePackage{fix-cm}
\documentclass[smallextended]{svjour3}
\journalname{Quantum Information Processing}

\usepackage{graphics}
\usepackage{graphicx}
\usepackage{url}
\usepackage{amssymb}
\usepackage{bbold}
\usepackage{a4wide}
\usepackage{enumitem}

\begin{document}

\setlength{\parindent}{0mm}
\thispagestyle{empty}

\begin{center}
{\Large\bf Security proof for\\ Round Robin Differential Phase Shift QKD}\\
\vskip1mm
Daan Leermakers and Boris \v{S}kori\'{c}
\\ TU Eindhoven\\ {\tt d.leermakers.1@tue.nl}, {\tt b.skoric@tue.nl}
\end{center}

{\it 
We give a security proof of the `Round Robin Differential Phase Shift' 
Quantum Key Distribution scheme,
and we give a tight bound on the required amount of privacy amplification.
Our proof consists of the following steps.
We construct an EPR variant of the scheme. 
We identify Eve's optimal way of
coupling an ancilla to an EPR qudit pair under the constraint that the bit error rate 
between Alice and Bob should not exceed a value~$\qb$.
As a function of $\qb$
we derive, for finite key size, the trace distance between the real state and
a state in which no leakage exists.
For asymptotic key size we obtain a bound on the trace distance
by computing the von Neumann entropy.
Our asymptotic result for the privacy amplification is sharper than
existing bounds.

}

\section{Introduction}

\subsection{Quantum Key Distribution and the RRDPS scheme}

Quantum-physical information processing is different from classical 
information processing in several remarkable ways.
Performing a measurement on an unknown quantum state typically destroys information; 
It is impossible to clone an unknown state by unitary evolution~\cite{WoottersZurek};
Quantum entanglement is a form of correlation between subsystems that does not exist
in classical physics.
Numerous ways have been devised to exploit these quantum properties
for security purposes \cite{BroadbentSchaffner2016}.
By far the most popular and well studied type of protocol is Quantum Key Distribution (QKD).
QKD was first proposed in a famous paper by Bennett and Brassard in 1984 \cite{BB84}.
Given that Alice and Bob have a way to authenticate classical messages to each other (typically a short key),
and that there is a quantum channel from Alice to Bob,
QKD allows them to create a random key of arbitrary length about which Eve knows practically nothing.
BB84 works with two conjugate bases in a two-dimensional Hilbert space.
Many QKD variants have since been described in the literature \cite{Ekert91,Bennett1992,Ralph1999,Hillery2000,GottPres2001,IWY2002},
using e.g.\,different sets of qubit states, EPR pairs, qudits instead of qubits, or continuous variables.
Furthermore, various proof techniques have been developed \cite{Bruss1998,SP2000,RGK2005,KGR2005}.

In 2014, Sasaki, Yamamoto and Koashi introduced {\em Round-Robin Differential Phase-Shift} (RRDPS) \cite{SYK2014},
a QKD scheme based on $d$-dimenional qudits. 
It has the advantage that it is very noise resilient while being easy to implement using photon pulse trains
and interference measurements.
One of the interesting aspects of RRDPS is that it is possible to omit the monitoring of signal disturbance.
Even at high disturbance, Eve can obtain little information $I_{\rm AE}$ about Alice's secret bit.
The value of $I_{\rm AE}$ determines how much privacy amplification is needed.
As a result of this, the maximum possible QKD rate 
(the number of actual key bits conveyed per quantum state)
is $1-h(\qb)-I_{\rm AE}$, where $h$ is the binary entropy function
and $\qb$ the bit error rate.\footnote{
Monitoring of signal disturbance induces a small penalty on the QKD rate.
However, the number of qubits that needs to be discarded is only logarithmic
in the length of the derived key \cite{Hoi-Kwong2005}
and hence we will ignore the penalty.
}

\subsection{Prior work on the security of RRDPS}
\label{sec:previoussec}

The security of RRDPS has been discussed in a number of papers \cite{SYK2014,ZYCM2015,SK2017,Sko2017}.
The original RRDPS paper gives an upper bound for asymptotic key length,
\be
	I_{\rm AE}\leq h(\frac1{d-1})
\label{IAESYK}
\ee
(Eq.\,5 in \cite{SYK2014} with photon number set to~1).
The security analysis in \cite{SYK2014} is based on an entropic inequality for non-commuting measurements.
There are two issues with this analysis.
First, the proof is not written out in detail.
Second, it is not known how tight the bound is.

Ref.\,\cite{ZYCM2015} follows \cite{SYK2014} and does a more accurate computation of phase error rate, tightening the
$1/(d-1)$ in (\ref{IAESYK}) to $1/d$.
In \cite{SK2017} Sasaki and Koashi add $\qb$-dependence to their analysis and claim a bound
\be
	I_{\rm AE}\leq h(\frac{2\qb}{d-2}) \quad\quad\mbox{for }\qb\leq \frac12\cdot\frac{d-2}{d-1}
\label{SKwithbeta}
\ee
and $I_{\rm AE}\leq h(\fr1{d-1})$ for $\qb\in[\fr12\cdot\fr{d-2}{d-1},\fr12]$.
The analysis in \cite{Sko2017} considers only intercept-resend attacks, and hence puts a {\em lower bound}
on Eve's potential knowledge,
$I_{\rm AE}\geq 1-h(\fr12+\fr1d)=\cO(1/d^2)$.\footnote{
Ref.\,\cite{Sko2017} gives a min-entropy of $-\log(\fr12+\fr1d)$, which translates to Shannon entropy
$h(\fr12+\fr1d)$.
}

\subsection{Contributions and outline}

In this paper we give a security proof 
of RRDPS in the case of coherent attacks. 
We give a bound on the required amount of privacy amplification.
We adopt a proof technique inspired by \cite{SP2000}, \cite{KGR2005} and \cite{Bruss1998}.
We consider the case where Alice and Bob {\em do monitor the channel} (i.e. they are able to tune
the amount of privacy amplification (PA) as a function of the observed bit error rate) as well as the saturated regime
where the leakage does not depend on the amount of noise.

\begin{itemize}[leftmargin=4mm]
\item
We show that the RRDPS protocol is equivalent to a protocol that contains an additional
randomisation step by Alice and Bob. The randomisation consists of phase flips and a permutation
of the basis states. 
We construct an EPR variant of RRDPS-with-randomisation; it is equivalent to RRDPS if Alice
creates the EPR pair and immediately does her measurement.\footnote{
This is similar to the Shor-Preskill technique \cite{SP2000}. 
}
The effect of the randomisation is that Alice and Bob's entangled state after Eve's attack
on the EPR pair is symmetrised
and can be described using just four real degrees of freedom. 
\item
We identify Eve's optimal way of
coupling an ancilla to an EPR qudit pair under the constraint that the bit error rate 
between Alice and Bob does not exceed some value~$\qb$.
\item
We consider an attack where Eve applies the above coupling to each EPR qudit-pair individually.
We compute an upper bound on the statistical distance of the full QKD key (after PA) from uniformity, conditioned on Eve's ancilla states.
From this we derive how much privacy amplification is needed.
The result does not depend on the way in which Eve uses her ancillas, i.e. she may apply a postponed coherent measurement 
on the whole system of ancillas.
\item
We compute the von Neumann mutual information between one ancilla state and Alice's secret bit.
This provides a bound on the leakage in the asymptotic (long key) regime \cite{RGK2005}.
Our result is sharper than \cite{SYK2014}.
\item
We provide a number of additional results by way of supplementary information.
(i) We show that Eve's ancilla coupling can be written as a unitary operation on the Bob-Eve system.
This means that the attack can be executed even if Eve has no access to Alice's qudit; this is important especially 
in the reduction from the EPR version to the original RRDPS.
(ii) We compute the min-entropy of one secret bit given the corresponding ancilla.
(iii) We compute the accessible information (mutual Shannon entropy) of one secret bit given 
the corresponding ancilla.
The min-entropy and accessible-information results are relevant for collective attacks.
\end{itemize}


In Section~\ref{sec:prelim} we introduce notation and briefly summarise 
the RRDPS scheme, the attacker model, and
extraction of classical information from (mixed) quantum states.
Section~\ref{sec:mainresult} states the main result: the amount of privacy amplification needed for 
RRDPS to be secure,
(i) at finite key length and (ii) asymptotically.
The remainder of the paper builds towards the proof of these results, and provides 
supplementary information about the leakage in terms of min-entropy loss and accessible (Shannon) information.

In Section~\ref{sec:EPR} we show that the randomisation step does not modify RRDPS, 
and we introduce the EPR version of the protocol. 
In Section~\ref{sec:noiseconstraint} we impose the constraint that Eve's actions 
must not cause a bit error rate higher than~$\qb$, and determine which mixed states of the Alice-Bob system
are still allowed. There are only two scalar degrees of freedom left, which we denote as $\mu$ and~$V$.
In Section~\ref{sec:purification} we do the purification of the Alice-Bob mixed state,
thus obtaining an expression for the state of Eve's ancilla.
Although the ancilla space has dimension $d^2$, we show that only a four-dimensional subspace
is relevant for the analysis.
In Section~\ref{sec:statdist} we 
prove the non-asymptotic main result by deriving an upper bound on
the statistical distance between the distribution of the QKD key
and the uniform distribution, conditioned on Eve's ancillas.
In Section~\ref{sec:vonNeumann} we prove the asymptotic result by
computing Eve's knowledge in terms of von Neumann entropy. 
In Section~\ref{sec:collective} we study collective attacks.
Section~\ref{sec:discussion} compares our results to previous bounds.

\section{Preliminaries}
\label{sec:prelim}

\subsection{Notation and terminology}
\label{sec:notation}
Classical Random Variables (RVs) are denoted with capital letters, and their realisations
with lowercase letters. The probability that a RV $X$ takes value
$x$ is written as $\pr[X=x]$.
The expectation with respect to RV $X$ is denoted as 
$\EE_x f(x)=\sum_{x\in\cX}\pr[X=x]f(x)$. The constrained sum
$\sum_{t,t':t\neq t'}$ is abbreviated as $\sum_{[tt']}$ and $\EE_{u,v:u\neq v}$ as $\EE_{[uv]}$.
The Shannon entropy of $X$ is written as $\sH(X)$.
Sets are denoted in calligraphic font. 
The notation `$\log$' stands for the logarithm with base~2.
The min-entropy of $X\in\cX$ is $\Hmin(X)=-\log \max_{x\in\cX}\pr[X=x]$,
and the conditional min-entropy is
$\Hmin(X|Y)=-\log\EE_y \max_{x\in\cX}\pr[X=x|Y=y]$.
The notation $h$ stands for the binary entropy function $h(p)=p\log\fr1p+(1-p)\log\fr1{1-p}$.
Bitwise XOR of binary strings is written as `$\oplus$'.
The Kronecker delta is denoted as $\qd_{ab}$.
For quantum states we use Dirac notation.
The notation `tr' stands for trace.
The Hermitian conjugate of an operator $A$ is written as~$A\dagg$.
When $A$ is a complicated expression, we sometimes write $(A+{\rm h.c.})$
instead of $A+A\dagg$.
The complex conjugate of $z$ is denoted as $z^*$.
We use the Positive Operator Valued Measure (POVM) formalism.
A POVM $\cM$ consists of positive semidefinite operators,
$\cM=(M_x)_{x\in\cX}$, $M_x\geq 0$, and satisfies the condition $\sum_x M_x=\one$. 
The trace norm of $A$ is $\| A\|_1=\tr \sqrt{A\dagg A}$.
The trace distance between matrices $\qr$ and $\qs$ is denoted as 
$D(\qr,\qs)=\frac{1}{2} \left\|\rho -\sigma \right\|_1$;
it is a generalisation of the statistical distance
and represents the maximum possible advantage one can have in distinguishing
$\qr$ from~$\qs$.

Consider a uniform classical variable $X$ and a mixed state $\qr(X)$
that depends on~$X$. 
The combined quantum-classical state is $\EE_x \ket x\bra x\otimes\qr(x)$.
The statistical distance between $X$ and a uniform variable given $\qr(X)$
is a measure of the security of $X$ given $\qr$.
This distance is given by \cite{RennerKonig2005} 
\be
	\cD(X|\qr(X))\isdef D\big( \EE_x \ket x\bra x\otimes\qr(x),\;
	\EE_x \ket x\bra x\otimes\EE_{x'}\qr(x') \big),
\label{defDXgivenrhoX}
\ee
i.e. the distance between the true quantum-classical state and 
a state in which the quantum state is decoupled from~$X$.
The term Privacy Amplification is abbreviated as PA.

\subsection{(Min-)entropy of a classical variable given a quantum state}
\label{sec:prelimHmin}
The notation $\cM(\qr)$ stands for the classical RV
resulting when $\cM$ is applied to mixed state~$\qr$.
Consider a bipartite system `AB' where the `A' part is classical, i.e.\,the state is of the form
$\qr^{\rm AB}=\EE_{x\in\cX}\ket x\bra x\otimes \qr_x$
with the $\ket x$ forming an orthonormal basis.
The min-entropy of the classical RV $X$ given part `B' of the system is \cite{KRS2009}
\be
	\Hmin(X|\qr_X)=-\log\max_\cM \EE_{x\in\cX}\tr [M_x \qr_x].
\label{defHmin}
\ee
Here $\cM=(M_x)_{x\in\cX}$ denotes a POVM.
Let $\qL\isdef\sum_x \qr_x M_x$.
If a POVM can be found that satisfies the condition\footnote{
Ref.\,\cite{Holevo1973} specifies a second condition, namely $\qL\dagg=\qL$.
However, the hermiticity of $\qL$ already follows from the condition (\ref{Holevotest}).} 
\cite{Holevo1973}
\be
	\forall_{x\in\cX}:\; \qL-\qr_x\geq 0,
\label{Holevotest}
\ee
then there can be no better POVM for guessing $X$ (but equally good POVMs may exist).
For states that also depend on a classical RV $Y\in\cY$, the min-entropy of $X$ 
given the quantum state and $Y$ is
\be
	\Hmin(X|Y,\qr_X(Y))=-\log\EE_{y\in\cY}\max_\cM \EE_{x\in\cX}\tr [M_x \qr_x(y)].
\label{defHmincondY}
\ee
A simpler expression is obtained when $X$ is a binary variable.
Let $X\in\bits$.

Then
\bea
	&\hspace{-4.5mm}X&\hspace{-4.5mm}\sim(p_0,p_1): \nn\\
	&{\sf H}_{\rm min}&(X|Y,\qr_X(Y)) = 
	-\log\left(\frac12+\frac12\EE_y\tr\Big\|p_0\qr_0(y)-p_1\qr_1(y)
	\Big\|_1\right).
\label{binaryHmin}
\eea
This generalizes in a straightforward manner for states that depend on multiple classical RVs.
The Shannon entropy of a classical variable given a measurement on a quantum state is given by
\be
	\sH(X|\qr_X)=\min_\cM\sH(X|\cM(\qr_X)).
\label{ruleShannonmin}
\ee
The `accessible information' is defined as the mutual information $\sH(X)-\sH(X|\qr_X)$.
In contrast to the min-entropy case,
there is no simple test analogous to (\ref{Holevotest}) which tells you whether a local minimum
in (\ref{ruleShannonmin}) is a global minimum.

\subsection{The RRDPS scheme in a nutshell}
\label{sec:RRDPS}
The dimension of the qudit space is~$d$. The basis states\footnote{
The physical implementation \cite{SYK2014} is a {\em pulse train}: a photon is split into $d$
coherent pieces which are released at different, equally spaced, points in time.}
are denoted as $\ket t$, with time indices $t\in\{0,\ldots,d-1\}$.
Whenever we use notation ``$t_1+t_2$'' it should be understood that the addition of time indices
is modulo~$d$.
The RRDPS scheme consists of the following steps.
\begin{enumerate}[leftmargin=5mm]
\item
Alice generates a random bitstring $a\in\bits^d$. 
She prepares the single-photon state 
\be
	\ket{\mu_a}\isdef \frac1{\sqrt d}\sum_{t=0}^{d-1}(-1)^{a_t}\ket{t}
\ee
and sends it to Bob.
\item
Bob chooses a random integer $r\in\{1,\ldots,d-1\}$.
Bob performs a POVM measurement $\cM^{(r)}$ described by a set of $2d$ operators
$(M^{(r)}_{ks})_{k\in\{0,\ldots,d-1\},s\in\bits}$,
\bea
\label{defPOVMM}
	M^{(r)}_{ks} = \frac12\ket{\qJ^{(r)}_{ks}}\bra{\qJ^{(r)}_{ks}}
	&& \quad\quad\quad
	\ket{\qJ^{(r)}_{ks}} = \frac{\ket{ k}+(-1)^s\ket{k+r}}{\sqrt 2}.
\eea
The result of the measurement $\cM^{(r)}$ on $\ket{\mu_a}$
is an random integer $k\in\{0,\ldots,d-1\}$
and a bit $s=a_k\oplus a_{k+r}$.\footnote{
The phase $(-1)^{a_k\oplus a_{k+r}}$ is the phase of the field oscillation 
in the $(k+r)$'th pulse relative to the $k$'th.
The measurement $\cM^{(r)}$ is an interference measurement where one path is delayed by $r$ time units.
}
\item
Bob announces $k$ and~$r$ over a public but authenticated channel.
Alice computes $s=a_k\oplus a_{k+r}$.
Alice and Bob now have a shared secret bit~$s$.
\end{enumerate}
This procedure is repeated multiple times. 

To detect eavesdropping, Alice and Bob can compare a randomly selected fraction of their secret bits. 
If this comparison is not performed, 
Alice and Bob have to assume that Eve learns as much as when causing bit error rate $\qb=\frac{1}{2}$. 
This mode of operation (without monitoring) was proposed in the original RRDPS paper \cite{SYK2014}.

Finally,
on the remaining bits Alice and Bob carry out
the standard procedures of information reconciliation and privacy amplification.

The security of RRDPS is intuitively understood as follows.
A measurement in a $d$-dimensional space cannot extract more than $\log d$ bits of information.
The state $\ket{\mu_a}$, however, contains $d-1$ pieces of information,
which is a lot more than $\log d$. 
Eve can learn only a fraction of the string $a$ embedded in the qudit.
Furthermore, what information she has is of limited use, because she cannot force Bob 
to select specific phases.
(i) She cannot force Bob to choose a specific $r$ value.
(ii) Even if she feeds Bob a state of the form $\ket{\qJ^{(r)}_{\ell u}  }$,
where $r$ accidentally equals Bob's $r$,
then there is a 
$\fr12$ probability that Bob's measurement $\cM^{(r)}$ yields $k\neq\ell$ {\em with random~$s$}.

\subsection{Attacker model; channel monitoring}
\label{sec:attack}

There is a quantum channel from Alice to Bob.
There is an authenticated but non-confidential classical channel between Alice and Bob.
We allow Eve to attack individual qudit positions in any way allowed by the laws of quantum physics, 
e.g.\,using unbounded quantum memory, entanglement, lossless operations, arbitrary POVMs, arbitrary unitary operators etc. 
All bit errors observed by Alice and Bob are assumed to be caused by Eve. 
Eve cannot influence the random choices of Alice and Bob, nor the state of their (measurement) devices.
There are no side channels.
This is the standard attacker model for
quantum-cryptographic schemes.
We consider the following channel monitoring technique.
Alice and Bob test the bit error rate {\em for each 
combination $(a,k)$ separately},
demanding that for each $(a,k)$ the observed bit error rate does not exceed~$\tilde\qb<\qb$.\footnote{
The gap between $\tilde\qb$ and $\qb$ must be properly chosen as a function of the number of samples and
the required confidence level.
}
Furthermore they test if $k$ is uniform for every~$a$.
Since Eve has no control over $r$, 
passing these tests implies that for all
$(a,k,r)$ the bit error probability does not exceed~$\qb$ with overwhelming probability.\footnote{
Any statistical uncertainty about the uniformity of $k$ due to finite sample size 
can be represented as a statistical distance $\qd_{\rm sampl}$ 
between the real state and the state that satisfies the constraints.
The parameter $\qd_{\rm sampl}$ will then appear, via the triangle inequality,
as an additional term in the expression for the trace distance in Theorems \ref{theorem:mainstatdist}
and~\ref{theorem:mainvonNeumann}.
}

The number of `sacrificed' qudits required to implement all the tests on the bit error rate
is of order $2^d\cdot d\cdot \log\qk$, where $\qk$ is the
length of the final key \cite{Hoi-Kwong2005}.
We will assume that $n$ is chosen sufficiently large to ensure $d 2^d\log \qk\ll n$.

We will analyze an attack in which Eve couples an ancilla to each EPR pair individually {\em in the same way},
i.e. causing the same bit error probability~($\qb$).
This looks like a serious restriction on Eve. However, it will turn out (Section~\ref{sec:statdist})
that the leakage is a concave function of~$\qb$, which means that it is sub-optimal for Eve to 
use different ways of coupling for different EPR pairs.

We will see that the leakage becomes constant when $\qb$ reaches a saturation point.
If Alice and Bob are willing to tolerate such a noise level, then channel monitoring is no longer necessary
for determining the leakage;
they just assume that the maximum possible leakage occurs. 
(Monitoring is still necessary to determine which error-correcting code should be applied.)

Note that for large $d$ it becomes impractical to determine the bit error rate for each
combination $(a,k)$ individually due to the exponential factor $2^d$; the saturation value of the leakage should be assumed.

\section{Main results}
\label{sec:mainresult}

Our first result is a non-asymptotic bound on the
secrecy of the QKD key.


\begin{theorem}
\label{theorem:mainstatdist}
Let ${\bf r}=(r_1,\ldots,r_n)$ be the values of the parameter $r$ in $n$ rounds of RRDPS,
and similarly ${\bf k}=(k_1,\ldots,k_n)$. 
Let $z\in\bits^\ell$ be the QKD key derived from the $n$ rounds.
Let $u$ be the (public) random seed used in the privacy amplification.
Let $\qo(z,u,{\bf r},{\bf k})$ be the joint state of Eve's $n$ ancillas.
The distance of $Z$ from uniformity, given all Eve's available information, classical and quantum,
can be bounded as
\bea
	\cD\big(Z|U,{\bf r},{\bf k},\qo(Z,U,{\bf r},{\bf k})\big) 
	< 
	\frac{1}{2} \sqrt{2^{ \ell-n(1-2 \log T) }}
\eea
where $T$ is given by
\bea
	\qb\leq \qb_* &:& \quad 
	T = 2\qb+\sqrt{1-2\qb}\Big[\sqrt{1-2\qb\frac{d-1}{d-2}}+\frac{\sqrt{2\qb}}{\sqrt{d-2}}\Big]
	\\
	\qb\geq\qb_* &:& \quad 
	T=2\qb_*+\sqrt{1-2\qb_*}\Big[\sqrt{1-2\qb_*\frac{d-1}{d-2}}+\frac{\sqrt{2\qb_*}}{\sqrt{d-2}}\Big]
\eea
and $\qb_*$ is a saturation value that depends on $d$ as
\be
	\qb_* = \frac{x_d/2}{1+x_d},
\label{bstard1}
\ee
where $x_d$ is the solution on $(0,1)$ of the equation 
\be
	(1-\frac{x}{d-2})^{\frac12}+ (1+\frac1{d-2})(1-\frac{x}{d-2})^{-\frac12}
	+\frac1{\sqrt{d-2}}(\sqrt x-\frac1{\sqrt x})-2=0.
\label{bstarcondition1}
\ee
\end{theorem}

For asymptotically large $n$, it has been shown \cite{RennerKonig2005}, 
using the properties of smooth R\'{e}nyi entropies, that
$
	\cD(Z|\cdots) \leq 
	\frac{1}{2} \sqrt{2^{ \ell - n (1 - I_{\rm AE})}},
$
where $I_{\rm AE}$ is the single-qudit
von Neumann information leakage $S(E)-S(E|S')$. Here `E' stands for Eve's ancilla state
and $S'$ is Alice's secret bit.

Our second result is a computation of the von Neumann leakage $I_{\rm AE}$ for RRDPS.

\begin{theorem}
\label{theorem:mainvonNeumann}
The information leakage about the secret bit S' given R, K and Eve's quantum state, in terms of von Neumann entropy, is given by:
\bea
	\qb\leq \qb_0 &:& \quad
	I_{\rm AE}= (1-2\qb)h(\frac1{d-2}\cdot\frac{2\qb}{1-2\qb})
	\\
	\qb\geq \qb_0 &:& \quad
	I_{\rm AE}=(1-2\qb_0)h(\frac1{d-2}\cdot\frac{2\qb_0}{1-2\qb_0}).
\eea
Here $\qb_0$ is a saturation value (different from $\qb_*$) given by
\be
	\qb_0=\frac12\Big[1+\frac1{(d-2)(1-y_d)}\Big]^{-1}
\ee
where $y_d$ is the unique positive root of the polynomial $y^{d-1}+y-1$. 
\end{theorem}

The theorems are proven in Sections \ref{sec:statdist} and \ref{sec:vonNeumann}.
The formulation of the security in terms of statistical distance ensures that 
the results are Universally Composable.
In Section~\ref{sec:vonNeumann} we will see that Theorem~\ref{theorem:mainvonNeumann}
is sharper than (\ref{SKwithbeta}) and hence allows for a higher QKD rate $\ell/n$.

\section{Symmetrised EPR version of the protocol}
\label{sec:EPR}

\subsection{RRDPS is equivalent to RRDPS with random permutations}
\label{sec:withperm}

We show that inserting a symmetrisation step into
RRDPS does not affect the protocol. More specifically, the following 
protocol is equivalent to RRDPS. 
\begin{enumerate}
\item[{\bf S1}]
Alice picks a random $a\in\bits^d$ and a random permutation~$\pi$.
She prepares $\ket{\mu_a}=\fr1{\sqrt d}\sum_t (-1)^{a_t}\ket t$.
\item[{\bf S2}]
Alice performs the permutation $\pi$ on the state $\ket{\mu_a}$.
She sends the result to Bob. After pausing for a while, she sends $\pi$ to Bob.
\item[{\bf S3}]
Eve does something with the state, without knowing~$\pi$. Then she sends the result to Bob.
\item[{\bf S4}]
Bob receives a state and stores it until he receives~$\pi$.
Bob applies $\pi^{-1}$ to the state.
\item[{\bf S5}]
Bob picks a random $r\in\{1,\ldots,d-1\}$ and does the $\cM^{(r)}$ POVM.
The result is an index $k\in\{0,\ldots,d-1\}$ and a bit $s=a_k\oplus a_{k+r}$.
He computes $\ell=k+r\mod d$.
He announces $k,\ell$.
\item[{\bf S6}] 
Alice computes $s'=a_k\oplus a_\ell$.
\end{enumerate}
The equivalence is shown as follows.
After step S2, the state is $\fr1{\sqrt d}\sum_t (-1)^{a_t}\ket{\pi(t)}$

$=\fr1{\sqrt d}\sum_\qt (-1)^{a_{\pi^{-1}\qt}}\ket \qt$
$=\ket{\mu_{\pi^{-1}(a)}}$. Hence Alice's process \{state preparation followed by~$\pi$\}
can be replaced by \{acting with $\pi^{-1}$ on $a$ followed by state preparation\}.
Similarly, Bob's process \{apply $\pi^{-1}$ to state; pick random $r$; do $\cM^{(r)}$; send $k,\ell$\}
has exactly the same effect as \{pick random $r$; do $\cM^{(r)}$; apply $\pi$ to $k,l$;
send $\pi(k),\pi(\ell)$\}.
Next, Bob's computation of $\pi(k),\pi(\ell)$ can be moved to Alice.
Then, Alice's actions \{pick random $a$; send $\pi^{-1}(a)$ to state preparation;
send $a$ to step S6\} can be replaced by \{pick random $a'$; send $a'$ to state preparation;
send $\pi(a)$ to step S6\}.
Finally, in step S6 we use $\pi(a)_{\pi(k)}=a_k$ and $\pi(a)_{\pi(\ell)}=a_\ell$.

{\em Remark}.
In step S3 it is crucial that Eve does not know $\pi$ at the moment of her manipulation of the state.
This will allow us to derive a symmetrised form of the density matrix
in Section~\ref{sec:toEPR}.

\subsection{RRDPS is equivalent to RRDPS with random phase flips}
\label{sec:withflips}

Analogous with Section~\ref{sec:withperm}, it can be seen that adding an extra phase-flipping step
to RRDPS does not affect RRDPS.
Consider the following protocol.
\begin{enumerate}
\item[{\bf F1}]
Alice picks a random $a\in\bits^d$ and a random $c\in\bits^d$.
She prepares $\ket{\mu_a}=\fr1{\sqrt d}\sum_t (-1)^{a_t}\ket t$.
\item[{\bf F2}]
Alice performs the phase flips on the state $\ket{\mu_a}$,
according to the rule $\ket t\to(-1)^{c_t}\ket t$ for basis states.
She sends the result to Bob. After pausing for a while, she sends $c$ to Bob.
\item[{\bf F3}]
Eve does something with the state, without knowing~$c$. Then she sends the result to Bob.
\item[{\bf F4}]
Bob receives a state and stores it until he receives~$c$.
Bob applies phase flips $c$ to the state.
\item[{\bf F5}]
Bob picks a random $r\in\{1,\ldots,d-1\}$ and does the $\cM^{(r)}$ POVM.
The result is an index $k\in\{0,\ldots,d-1\}$ and a bit $s=a_k\oplus a_{k+r}$.
He computes $\ell=k+r\mod d$.
He announces $k,\ell$.
\item[{\bf F6}] 
Alice computes $s'=a_k\oplus a_\ell$.
\end{enumerate}
The equivalence to RRDPS is seen as follows.
After step F2 the state is $\ket{\mu_{a\oplus c}}$. Hence Alice's process 
\{pick random $a$; prepare state; flip with $c$\}
is equivalent to
\{pick random $a$; flip with $c$; prepare state\}.
Similarly, Bob's process
\{flip with $c$; pick random r; do $\cM^{(r)}$\}
is equivalent to 
\{pick random r; do $\cM^{(r)}$; change $s$ to $s\oplus c_k\oplus c_\ell$ \}.
This holds because in the first case Bob obtains 
$s=(a\oplus c)_k\oplus(a\oplus c)_\ell=(a_k\oplus a_\ell)\oplus c_k\oplus c_\ell$.
Furthermore, Alice's steps
\{pick random a; send $a$ to computation of $s'$ and flipped $a$ to state preparation\}
are equivalent to
\{pick random $a'$; send flipped $a$ to computation of $s'$ and $a'$ to state preparation\}.
The final effect of these transformations of the `F' protocol is that 
(i) there is no physical phase flipping
at all,
(ii) Bob needs no quantum memory,
and (iii)
Alice and Bob both obtain a secret bit $(a_k\oplus a_\ell)\oplus c_k\oplus c_\ell$;
though not equal to $a_k\oplus a_\ell$, it is statistically the same.

\subsection{EPR version}
\label{sec:toEPR}

We introduce a protocol based on EPR pairs that is equivalent to the combined `S' and `F' protocols, 
and hence also equivalent to RRDPS.

\begin{enumerate}
\item[{\bf E1}]
A maximally entangled two-qudit state is prepared.
\be
	\ket{\qa_0}\isdef\frac1{\sqrt d}\sum_{t=0}^{d-1}\ket{tt}.
\label{defGamma}
\ee
One qudit (`A') is intended for Alice, and one (`B') for Bob.
\item[{\bf E2}]
Eve does something with the EPR pair.
Then Alice and Bob each receive their own qudit.
\item[{\bf E3}]
Alice and Bob pick a random permutation $\pi$. They both apply $\pi$
to their own qudit. Then they forget~$\pi$.
\item[{\bf E4}]
Alice and Bob pick a random string $c\in\bits^d$. They both apply 
phase flips $\ket t\to(-1)^{c_t}\ket t$
to their own qudit. Then they forget~$c$.
\item[{\bf E5}]
Alice performs a POVM $\cQ=(Q_z)_{z\in\bits^d}$ on her own qudit, where
\be
	Q_z=\frac{d}{2^d}\ket{\mu_z}\bra{\mu_z}.
\label{defPOVMQ}
\ee
This results in a measured string~$a\in\bits^d$.
\item[{\bf E6}]
Bob picks a random integer $r\in\{1,\ldots,d-1\}$ and performs the POVM
measurement $\cM^{(r)}$ on his qudit.
The result of the measurement is an integer $k\in\{0,\ldots,d-1\}$
and a bit~$s$. 
Bob computes $\ell=k+r\mod d$.
Bob announces $k,\ell$. 
\item[{\bf E7}]
Alice computes $s'=a_k\oplus a_\ell$.
\end{enumerate}

The equivalence to the protocol in Section~\ref{sec:withperm} is seen as follows.
First, let Alice be the origin of the EPR pair, and let her perform $\cQ$
as soon as she has created the EPR pair.
This process is equivalent to preparing a qudit state $\ket{\mu_a}$
with random~$a$.
The only difference is that the EPR protocol allows 
Eve to couple her ancilla to the AB system instead of only the B system.
Hence the EPR version {\em overestimates} Eve's power.
Security of the EPR version implies security of the original RRDPS.\footnote{
In Appendix~\ref{app:BE} it will turn out that Eve's optimal attack 
is achieved by acting on Bob's qudit only; hence the EPR version is fully equivalent to 
original RRDPS.
}
Furthermore, the permutations and phase flips in steps E3,E4 cancel out exactly like in
protocols `S' and `F'.

{\em Remark:}
The protocol equivalences is Sections~\ref{sec:withperm}--\ref{sec:toEPR}
can be nicely visualised using diagrammatic techniques \cite{AleksBob}.
We do not show the protocol diagrams in this paper.


\begin{lemma}
\label{lemma:AlicePOVM}
The hermitian matrices $Q_z$ as defined in (\ref{defPOVMQ}) form a POVM, i.e.
	$\sum_{z\in\bits^d}Q_z=\one$.
\end{lemma}
\underline{\it Proof:}\\
$\sum_z\ket{\mu_z}\bra{\mu_z}=$
$\sum_z\fr1d\sum_{t,t'=0}^{d-1}(-1)^{z_{t'}+z_{t}}\ket t\bra{t'}$
$=\fr1d\sum_{t,t'=0}^{d-1}\ket t\bra{t'}\sum_z(-1)^{z_{t'}+z_t}$.
\newline
Using $\sum_z(-1)^{z_{t'}+z_t}=2^d\qd_{tt'}$ we get
$\sum_z\ket{\mu_z}\bra{\mu_z}=\fr{2^d}d\sum_t\ket t\bra t=\fr{2^d}d\one$.
\hfill$\square$

Alice and Bob's measurements can be carried out in the opposite order.
It is not important whether $\cQ$ is practical or not; it is a theoretical construct which allows 
us to build an EPR version of RRDPS. 
 
\subsection{Effect of the random transforms: state symmetrisation}
\label{sec:symmetrisation}

Let $\qr^{\rm AB}$ denote the pure EPR state of Alice and Bob,
and let $\hat\qr^{\rm AB}$ be the mixed state of the AB system
after Eve's manipulation in step~E2.
We write
\be
	\hat\qr^{\rm AB}=\sum_{t,t',\qt,\qt'\in\{0,\ldots,d-1\}}\hat\qr^{tt'}_{\qt\qt'}\ket{t,t'}\bra{\qt,\qt'},
\label{genrhomixed}
\ee
with $\hat\qr^{\qt\qt'}_{tt'}=(\hat\qr^{tt'}_{\qt\qt'})^*$ and $\sum_{tt'}\hat\qr^{tt'}_{tt'}=1$.
The effect of step E3 is that the AB state
gets averaged over all permutations, i.e. we get the following mapping
\bea
	\hat\qr^{\rm AB} \mapsto \tilde\qr^{\rm AB} &\isdef &
	\frac1{d!}\sum_\pi\sum_{t,t',\qt,\qt'} \hat\qr^{\pi(t),\pi(t')}_{\pi(\qt),\pi(\qt')}\ket{t,t'}\bra{\qt,\qt'}
	\\ &\isdef & 
	\sum_{t,t',\qt,\qt'}\tilde\qr^{tt'}_{\qt\qt'} \ket{t,t'}\bra{\qt,\qt'}.
\eea
Here the parameters $\tilde\qr^{tt'}_{\qt\qt'}$ are invariant under simultaneous permutation of the
four indices, i.e. $\tilde\qr^{\pi(t),\pi(t')}_{\pi(\qt),\pi(\qt')}=\tilde\qr^{tt'}_{\qt\qt'}$ 
for all $\pi$,$t$,$t'$,$\qt$,$\qt'$.
The consequence is that $\tilde\qr^{\rm AB}$ contains only a few degrees of freedom, namely 
the constants $\tilde\qr^{ss}_{ss}$, 
$\tilde\qr^{ss}_{st}$, $\tilde\qr^{ss}_{ts}$,
$\tilde\qr^{ss}_{tt}$, $\tilde\qr^{st}_{st}$, $\tilde\qr^{st}_{ts}$,
$\tilde\qr^{ss}_{tu}$, $\tilde\qr^{st}_{su}$, $\tilde\qr^{ts}_{us}$, $\tilde\qr^{st}_{us}$,
$\tilde\qr^{st}_{uv}$,
where $s,t,u,v$ are mutually distinct.

Next, the random phase flips reduce the degrees of freedom even further.
Let $F_c$ be the phase flip operator.
\bea
	\bar\qr^{\rm AB} &\isdef& \EE_{c\in\bits^d} F_c\tilde\qr^{\rm AB}F_c\dagg
	\\ &=&
	\EE_c \sum_{tt'\qt\qt'}\tilde\qr^{tt'}_{\qt\qt'} (-1)^{c_t+c_{t'}+c_\qt+c_{\qt'}}
	\ket{t,t'}\bra{\qt,\qt'}
	\\ &=&
	\sum_{tt'\qt\qt'} \ket{t,t'}\bra{\qt,\qt'} 
	 \tilde\qr^{tt'}_{\qt\qt'} \EE_c (-1)^{c_t+c_{t'}+c_\qt+c_{\qt'}}
\label{EEcflips}
	\\ &\isdef&
	\sum_{tt'\qt\qt'} \ket{t,t'}\bra{\qt,\qt'}\bar\qr^{tt'}_{\qt\qt'}.
\eea
From (\ref{EEcflips}) we see that any time index that occurs an odd number of times
will be wiped out, i.e. $\EE_c (-1)^{c_t}=0$.
The only surviving degrees of freedom are the constants $\bar\qr^{ss}_{ss}$, $\bar\qr^{ss}_{tt}$,
$\bar\qr^{st}_{st}$ and $\bar\qr^{st}_{ts}$ (all with $t\neq s$ and arbitrary $s,t$).
Note that all four constants all real-valued.

\section{Imposing the noise constraint}
\label{sec:noiseconstraint}

The channel monitoring 
restricts the ways in which Eve can alter the AB state without being detected.
We will determine the most general allowed $\bar\qr^{\rm AB}$
that is compatible with bit error rate $\qb$ for all values of $(a,k,r,s)$.
(We will later see that it is optimal for Eve to cause the same bit error rate
in all rounds. This is due to the concavity of the leakage as a function of the error rate.)
We introduce the notation $P_{aks|r}=\pr[A=a,K=k,S=s|R=r]$.

 

\begin{lemma}
\label{Paksr}
Let Alice and Bob's bipartite state be $\bar\qr^{\rm AB}$,
and let them perform the measurements $\cQ$ and $\cM^{(r)}$ respectively.
At given $r$,
the joint probability of the outcomes $a,k,s$ is given by
\be
	P_{aks|r}=\frac1{4\cdot 2^d}\sum_{t\qt}(-1)^{a_t+a_\qt}
	[\bar\qr^{tk}_{\qt k}+\bar\qr^{t,k+r}_{\qt,k+r}+(-1)^s(\bar\qr^{tk}_{\qt,k+r}+\bar\qr^{t,k+r}_{\qt k})].
\label{noisePaksr}
\ee
\end{lemma}
\underline{\it Proof:}
$P_{aks|r}=\tr (Q_a\otimes M^{(r)}_{ks})\bar\qr^{\rm AB}$
\newline
$=\tr(\fr1{2^d}\sum_{\ell\ell'}(-1)^{a_\ell+a_{\ell'}}\ket\ell\bra{\ell'}\otimes
\frac12\frac{\ket k+(-1)^s\ket{k+r}}{\sqrt2} \frac{\bra k+(-1)^s\bra{k+r}}{\sqrt2})
\sum_{tt'\qt\qt'}\bar\qr^{tt'}_{\qt\qt'}\ket t\bra\qt\otimes\ket{t'}\bra{\qt'}$
\newline
$=\frac{1}{2^d 4}\sum_{tt'\qt\qt'}\bar\qr^{tt'}_{\qt\qt'}(-1)^{a_t+a_\qt}
[\qd_{t'k}+(-1)^s \qd_{t',k+r}] [\qd_{\qt'k}+(-1)^s \qd_{\qt',k+r}]$
\newline
$=\frac{1}{2^d 4}\sum_{t\qt}(-1)^{a_t+a_\qt}[\bar\qr^{tk}_{\qt k}+\bar\qr^{t,k+r}_{\qt,k+r}
+(-1)^s \bar\qr^{tk}_{\qt,k+r}+(-1)^s \bar\qr^{t,k+r}_{\qt k}]$.
\hfill$\square$

We now impose the constraint that 
the event $s\neq s'$ occurs with probability $\qb$ for all combinations $(a,k,r)$,
\be
	\forall_{aksr}:\quad
	P_{aks|r}^{\rm constr}=\frac1{2^d d}\Big[\qd_{s,a_k\oplus a_{k+r}}(1-\qb)+(1-\qd_{s,a_k\oplus a_{k+r}})\qb\Big].
\label{mainconstraint}
\ee

\begin{theorem}
\label{th:imposeconstraint}
The constraint (\ref{mainconstraint})
can only be satisfied by a density function of the form
\be
	\bar\qr^{\rm AB}=(1-2\qb-V)\ket{\qa_0}\bra{\qa_0} +V\frac1d\sum_{tt'}\ket{tt'}\bra{t' t}
	+(2\qb-\mu)\frac\one{d^2}+\mu\frac1d\sum_t\ket{tt}\bra{tt}
\label{reduced}
\ee
with $\mu,V\in\RR$. Written componentwise,
\be
	\bar\qr_{\qt\qt'}^{tt'}=\frac{1-2\qb-V}d\qd_{t't}\qd_{\qt'\qt}+\frac Vd\qd_{\qt t'}\qd_{\qt' t}
	+\frac{2\qb-\mu}{d^2}\qd_{\qt t}\qd_{\qt' t'}
	+\frac\mu d\qd_{t't}\qd_{\qt t}\qd_{\qt't}.
\ee
\end{theorem}
\underline{\it Proof:}
We rewrite the constraint (\ref{mainconstraint}) as
\be
	\forall_{aksr}:\quad P_{aks|r}^{\rm constr}=
	\frac1{2d 2^d}+\frac{\fr12-\qb}{2^d d}(-1)^s(-1)^{a_k\oplus a_{k+r}}.
\label{constrrewritten}
\ee
Then we use the fact that $\bar\qr^{\rm AB}$ depends only on four real-valued
constants, which we write as
$u\isdef \bar\qr^{ss}_{ss}$, $w\isdef \bar\qr^{st}_{st}$, 
$x\isdef\bar\qr^{ss}_{tt}$, $y\isdef\bar\qr^{st}_{ts}$ (with $s\neq t$ and arbitrary $s,t$).
In terms of these constants, the probability (\ref{noisePaksr}) is expressed as
\be
	P_{aks|r}=\frac{u+(d-1)w}{2\cdot 2^d} +\frac{x+y}{2\cdot 2^d}(-1)^s(-1)^{a_k\oplus a_{k+r}}.
\ee
Having $P_{aks|r}=P_{aks|r}^{\rm constr}$ 
requires setting
$u+(d-1)w=\fr1d$ and $(x+y)d=1-2\qb$.
The state $\bar\qr^{\rm AB}$ has the form
\be
	\bar\qr^{\rm AB}=u\sum_t\ket{tt}\bra{tt}+w\sum_{[tt']}\ket{tt'}\bra{tt'}
	+x\sum_{[t\qt]}\ket{tt}\bra{\qt\qt} + y\sum_{[tt']}\ket{tt'}\bra{t't}.
\label{genformbarqrAB}
\ee
We choose $x,w$ as the two independent degrees of freedom and re-parametrise them as
$x=(1-2\qb-V)/d$ and $w=(2\qb-\mu)/d^2$, where $\mu,V\in\RR$ are the new independent degrees of freedom. 
Substitution of $u=\fr1d-(d-1)w$, $y=(1-2\qb)/d-x$ and the re-parametrisation into 
(\ref{genformbarqrAB})
yields (\ref{reduced}).
\hfill$\square$

Theorem~\ref{th:imposeconstraint} shows that (at fixed $\qb$) there are only two degrees of freedom, $\mu$ and $V$,
in Eve's manipulation of the EPR pair.

\section{Purification}
\label{sec:purification}

According to the attacker model we have to assume that Eve has the purification 
of the state $\bar\qr^{\rm AB}$. The purification contains all information that exists outside the AB system. 

\subsection{The purified state and its properties}
\label{sec:purifiedprop}

We introduce the following notation,
\bea
	\ket{\qa_j} &\isdef& \frac1{\sqrt d}\sum_t e^{i\frac{2\pi}d jt}\ket{tt},
	\quad\quad j\in\{0,\ldots,d-1\}\\
	\ket{D_{tt'}^\pm}&\isdef& \frac{\ket{tt'}\pm\ket{t't}}{\sqrt2}
	\quad\quad t<t'.
\eea

\begin{lemma}
\label{lem:eigs}
The $\bar\qr^{\rm AB}$ given in (\ref{reduced})
has the following orthonormal eigensystem,
\bea
	\ket{\qa_0} && \mbox{with eigenvalue } \ql_0\isdef 
	\frac{2\qb-\mu}{d^2} +\frac{\mu+V}d+ 1-2\qb-V
	\nn\\
	\ket{\qa_j} \quad j\in\{1,\ldots,d-1\}
	&& \mbox{with eigenvalue } \ql_1\isdef\frac{2\qb-\mu}{d^2}+\frac{\mu+V}{d}.\\
	\ket{D_{tt'}^\pm}\quad (t<t') && \mbox{with eigenvalue } \ql_\pm\isdef\frac{2\qb-\mu}{d^2}\pm \frac Vd
	\nn
\eea
\end{lemma}
\underline{\it Proof:} 
The term proportional to $\one$ in (\ref{reduced}) yields a contribution $(2\qb-\mu)/d^2$ to each eigenvalue.
First we look at $\ket{\qa_j}$. We have $\inprod{\qa_0}{\qa_j}=\qd_{j0}$.
Furthermore $\inprod{t't}{\qa_j}=\qd_{t't}e^{i\frac{2\pi}{d}jt}/\sqrt d$,
which gives $(\sum_{tt'}\ket{tt'}\bra{t't})\ket{\qa_j}=\ket{\qa_j}$.
Similarly we have $(\sum_{t}\ket{tt}\bra{tt})\ket{\qa_j}=\ket{\qa_j}$.
Next we look at $\ket{D^\pm_{tt'}}$. We have $\inprod{\qa_0}{D^\pm_{tt'}}=0$ 
and $\inprod{uu}{D^\pm_{tt'}}=0$. Hence the $(1-2\qb-V)$-term and the $\mu$-term in (\ref{reduced})
yield zero when acting on $\ket{D^\pm_{tt'}}$.
Furthermore $\sum_{uu'}\ket{uu'}\inprod{u'u}{D^+_{tt'}}$
$=\sum_{uu'}\ket{uu'}\frac{\qd_{ut}\qd_{u't'}+\qd_{ut'}\qd_{u't}}{\sqrt2}$
$=\ket{D^+_{tt'}}$.
Similarly, $\sum_{uu'}\ket{uu'}\inprod{u'u}{D^-_{tt'}}$
$=\sum_{uu'}\ket{uu'}\frac{\qd_{ut}\qd_{u't'}-\qd_{ut'}\qd_{u't}}{\sqrt2}\sgn(u-u')$
$=-\ket{D^-_{tt'}}$.
\hfill$\square$\\

In diagonalised form the $\bar\qr^{\rm AB}$ is given by
\be
	\bar\qr^{\rm AB}=\ql_0\ket{\qa_0}\bra{\qa_0}+\ql_1\sum_{j=1}^{d-1}\ket{\qa_j}\bra{\qa_j}
	+\ql_+\sum_{tt':t<t'}\ket{D_{tt'}^+}\bra{D_{tt'}^+}
	+\ql_-\sum_{tt':t<t'}\ket{D_{tt'}^-}\bra{D_{tt'}^-}.
\ee
The purification is
\bea
	\ket{\qJ^{\rm ABE}}&=& \sqrt{\ql_0} \ket{\qa_0}\otimes \ket{E_0} 
	+\sqrt{\ql_1}\sum_{j=1}^{d-1}\ket{\qa_j}\otimes\ket{E_j}
	\nn\\ &&
	+\sqrt{\ql_+}\sum_{tt':t<t'}\ket{D_{tt'}^+}\otimes\ket{E_{tt'}^+}
	+\sqrt{\ql_-}\sum_{tt':t<t'}\ket{D_{tt'}^-}\otimes\ket{E_{tt'}^-}.
\label{PsiABE}
\eea
where we have introduced orthonormal basis states 
$\ket{E_j}$, $\ket{E_{tt'}^\pm}$ in Eve's Hilbert space. 
In Appendix~\ref{app:BE} we give more details on Eve's unitary operation.

\subsection{Eve's state}
\label{sec:Evestate}
Eve waits for Alice and Bob to perform their measurements and reveal $k$ and $r$. 
\begin{lemma}
\label{lemma:Evestate}
After Alice has measured $a\in\bits^d$ and Bob has measured $k\in\{0,\ldots,d-1\}$, $s\in\bits$,
Eve's state is given by
\be
	\qs^{rk}_{as}=\tr_{\!\rm AB} \Big[\ket{\qJ^{\rm ABE}}\bra{\qJ^{\rm ABE}}
	\frac{Q_a\otimes M^{(r)}_{ks}\otimes\one}{P_{aks|r}} \Big].
\label{Evestate}
\ee
\end{lemma}

\underline{\it Proof:}
The POVM elements $Q_a$ and $M^{(r)}_{ks}$ are proportional to projection operators.
Hence the tripartite ABE pure state after the measurement
is proportional to 
$(Q_a\otimes M^{(r)}_{ks}\otimes\one)\ket{\qJ^{\rm ABE}}$.
It is easily verified that the normalisation in (\ref{Evestate}) is correct:
taking the trace in E-space yields 
$\tr_{\!\rm AB}\tr_{\!\rm E}\ket{\qJ^{\rm ABE}}\bra{\qJ^{\rm ABE}}
Q_a\otimes M^{(r)}_{ks}\otimes\one$
$=\tr_{\!\rm AB}\;\bar\qr^{\rm AB}Q_a\otimes M^{(r)}_{ks}$
$=P_{aks|r}$.
\hfill$\square$

\begin{lemma}
\label{lemma:asumpartial}
It holds that
\bea
	\frac d{2^d} \hskip-5mm
	\sum_{\stackrel{a_0\cdots a_{d-1}}{{\rm without}\, a_k,a_{k+r}}} 
	\hskip-7mm
	\ket{\mu_a}\bra{\mu_a}
	&=& \frac14\one +\frac14(-1)^{a_k+a_{k+r}}\Big(\ket k\bra{k+r}+\ket{k+r}\bra k\Big)
	\\
	&=& M^{(r)}_{k,a_k\oplus a_{k+r}} + \frac14\sum_{t:\; t\neq k,k+r}\ket t\bra t.
\eea
\end{lemma}
\underline{\it Proof:}
We have $\ket{\mu_a}\bra{\mu_a}=\frac1d\one+\frac1d\sum_{[t\qt]}\ket t\bra\qt (-1)^{a_t+a_\qt}$.
Summation of the $\fr1d\one$ term is trivial and yields $2^{d-2}\cdot \frac1d\one$.
In the summation of the factor $(-1)^{a_t+a_\qt}$
in the second term, any summation $\sum_{a_t}(-1)^{a_t}$ yields zero.
The only nonzero contribution arises when $t=k,\qt=k+r$ or $t=k+r,\qt=k$; the a-summation then yields a factor $2^{d-2}$. 
\hfill$\square$

\begin{lemma}
\label{lemma:asumfull}
It holds that 
\be
	\EE_{a:a_k\oplus a_{k+r}=s'}\ket{\mu_a}\bra{\mu_a}=
	\frac\one d+(-1)^{s'}\frac{\ket k\bra{k+r}+\ket{k+r}\bra k}d.
\label{asumresult}
\ee
\end{lemma}
\underline{\it Proof:}
We have $\EE_{a:a_k\oplus a_{k+r}=s'}\ket{\mu_a}\bra{\mu_a}
=2^{-(d-1)}\sum_{a_k}\sum_{a_{k+r}}\qd_{a_k\oplus a_{k+r},s'}\cdot$\\
$\sum_{a\,{\rm without}\,a_k,a_{k+r}}\ket{\mu_a}\bra{\mu_a}$.
For the rightmost summation we use Lemma~\ref{lemma:asumpartial}.
Performing the $\sum_{a_k}$ and $\sum_{a_{k+r}}$ summations yields (\ref{asumresult}).
\hfill$\square$

Eve's task is to guess Alice's bit $s'=a_k\oplus a_{k+r}$ from the mixed state $\qs^{rk}_{as}$,
where Eve does not know $a$ and $s$.
We define
\be
	\qs^{rk}_{s'}=\EE_{s,a: a_k\oplus a_{k+r}=s'} [\qs^{rk}_{as}].
\ee
This represents Eve's ancilla state given some value of Alice's bit~$s'$.
%
Next we introduce notations that are useful for understanding the structure
of $\qs^{rk}_{s'}$.
We define, for $t,t'\in\{0,\ldots,d-1\}$, non-normalised
vectors $\ket{w_{tt'}}$ in Eve's Hilbert space as
\be
	\ket{w_{tt'}}\isdef\inprod{tt'}{\qJ^{\rm ABE}}.
\ee
Furthermore we define angles $\qa$ and $\qf$ as
\be
	\cos 2\qa \isdef \frac{\inprod{w_{kk}}{w_{k+r,k+r}}}{\inprod{w_{kk}}{w_{kk}}} 
	, \quad
	\cos2\qf \isdef \frac{\inprod{w_{k,k+r}}{w_{k+r ,k}}}{\inprod{w_{k,k+r}}{w_{k,k+r}}}
\ee
and vectors $\ket A, \ket B, \ket C, \ket D$
\bea
	\frac{\ket{w_{kk}}} {\sqrt{\inprod{w_{kk}}{w_{kk}}}}
	&=& \cos\qa\ket A+\sin\qa\ket B
	\\
	\frac{\ket{w_{k+r,k+r}}} {\sqrt{\inprod{w_{k+r,k+r}}{w_{k+r,k+r}}}} 
	&=& \cos\qa\ket A-\sin\qa\ket B
	\\
	\frac{\ket{w_{k,k+r}}} {\sqrt{\inprod{w_{k,k+r}}{w_{k,k+r}}}}
	&=&\cos\qf\ket C+\sin\qf\ket D
	\\
	\frac{\ket{w_{k+r, k}}} {\sqrt{\inprod{w_{k+r, k}}{w_{k+r, k}}}}
	&=&\cos\qf\ket C-\sin\qf\ket D.
\eea
The $\ket A$, $\ket B$, $\ket C$, $\ket D$ are mutually orthogonal,
and also orthogonal to any vector $\ket{w_{tt'}}$ ($t'\neq t$)
with $\{t,t'\}\neq\{k,k+r\}$.

\begin{theorem}
\label{th:sigmafull}
The eigenvalues of $\qs^{rk}_{s'}$ are given by
\bea
	\xi_0 &\isdef& \frac d2\cdot\frac{\ql_+ + \ql_-}2
	\\
	\xi_1 &\isdef& \fr d2(\ql_1+\ql_-) = \qb-\fr d2(\fr d2-1)(\ql_++\ql_-)
	\\
	\xi_2 &\isdef& \fr d2(\ql_1+2\frac{\ql_0-\ql_1}d+\ql_+) =1-\qb- \fr d2(\fr d2-1)(\ql_++\ql_-)
\eea
and the diagonal representation of $\qs^{rk}_{s'}$ is
\bea
	\qs^{rk}_{s'} &=& 
	\xi_0\sum_{{t\in\{0,\ldots,d-1\}}\atop{t\neq k,t\neq k+r}} 
	\Big(  \frac{\ket{w_{tk}}\bra{w_{tk}}}  {\inprod{w_{tk}}{w_{tk}}}
	+ \frac{\ket{w_{t,k+r}}\bra{w_{t,k+r}}} {\inprod{w_{t,k+r}}{w_{t,k+r}}}  \Big)
	\nn\\ &&
	+\xi_2
	\frac{[\sqrt{\xi_2-\fr d2\ql_+}\ket A+(-1)^{s'}\sqrt{\fr d2\ql_+}\ket C][\cdots]\dagg}{\xi_2}
	\nn\\ &&
	+\xi_1\frac{[\sqrt{\xi_1-\fr d2\ql_-}\ket B-(-1)^{s'}\sqrt{\fr d2\ql_-}\ket D][\cdots]\dagg}{\xi_1}
\label{sigmadiag}
\eea
\end{theorem}

\underline{\it Proof:}  
We have
\bea
\label{sigma_s'}
	\qs^{rk}_{s'}&=&\tr_{\!\rm AB}\ket{\qJ^{\rm ABE}}\bra{\qJ^{\rm ABE}}\EE_{a: a_k\oplus a_{k+r}=s'}Q_a\otimes
	\EE_{s|s'}\frac{M^{(r)}_{ks}}{P_{aks|r}}\otimes\one
	\nn\\ &=&
	d 2^d\; \tr_{\!\rm AB}\ket{\qJ^{\rm ABE}}\bra{\qJ^{\rm ABE}}[\EE_{a: a_k\oplus a_{k+r}=s'}Q_a]\otimes[\sum_s M^{(r)}_{ks}]
	\otimes\one.
	\label{sigmalong}
\eea
We use Lemma~\ref{lemma:asumfull} to evaluate the $\EE_a$ factor. We use $\sum_s M^{(r)}_{ks}=\fr12\ket k\bra k+\fr12\ket{k+r}\bra{k+r}$.
This allows us to write everything in terms of $\ket{w_{tt'}}$ states.
For $t=t'$ we have
\bea
	\ket{w_{tt}} &=& 
	\sqrt{\ql_0/d}\ket{E_0}+\sqrt{\ql_1/d} \sum_{j=1}^{d-1}(e^{i\frac{2\pi}d})^{jt}\ket{E_j}
	\\
	\inprod{w_{tt}}{w_{tt}} &=& \ql_1+\frac{\ql_0-\ql_1}d,
\eea
and for $t\neq t'$ we have
\bea
	\ket{w_{tt'}} &=& 
	\sqrt{\ql_+/2}\ket{E^+_{(tt')}}+\sgn(t'-t)\sqrt{\ql_-/2}\ket{E^-_{(tt')}}
	\\
	\inprod{w_{tt'}}{w_{tt'}} &=& (\ql_+ + \ql_-)/2.
\eea
The following properties hold ($t\neq t'$)
\bea
	\inprod{w_{tt}}{w_{tt'}}=0 &,&\quad \inprod{w_{tt}}{w_{t't}}=0 
	\\
	\inprod{w_{tt}}{w_{t't'}}=\frac{\ql_0-\ql_1}d &,&  \quad
	\inprod{w_{tt'}}{w_{t't}}=\frac{\ql_+ - \ql_-}2.
\eea
We get
\be
	\cos 2\qa =  1-\frac{d \ql_1}{\ql_0+(d-1)\ql_1}
	, \quad
	\cos2\qf = 1-\frac{2\ql_-}{\ql_++\ql_-}
\ee
After some tedious algebra the result (\ref{sigmadiag}) follows.
\hfill$\square$

Note that
the $\qs^{rk}_0$ and $\qs^{rk}_1$ have the same set of eigenvalues:
$2(d-2)$ times $\xi_0$, and once $\xi_1$ and $\xi_2$.


\begin{corollary}
\label{corol:sigmaav}
It holds that
\bea
	\frac{\qs^{rk}_0+\qs^{rk}_1}2 &=&
	\sum_{{t\in\{0,\ldots,d-1\}}\atop{t\neq k,t\neq k+r}} \xi_0\cdot
	\Big(  \frac{\ket{w_{tk}}\bra{w_{tk}}}  {\inprod{w_{tk}}{w_{tk}}}
	+ \frac{\ket{w_{t,k+r}}\bra{w_{t,k+r}}} {\inprod{w_{t,k+r}}{w_{t,k+r}}}  \Big)
	\nn\\ &&
	+(\xi_2-\fr d2\ql_+)\ket A\bra A+\fr d2\ql_+\ket C\bra C
	+(\xi_1-\fr d2\ql_-)\ket B\bra B+ \fr d2\ql_-\ket D\bra D.
	\nn
\eea
\end{corollary}
\underline{\it Proof:} Follows directly from Theorem~\ref{th:sigmafull} by
discarding the terms in (\ref{sigmadiag})
that contain $(-1)^{s'}$ (the AC and BD crossterms). 
\hfill$\square$

\begin{corollary}
\label{corol:sigdiff2}
The difference between $\qs^{rk}_0$ and $\qs^{rk}_1$ can be written as
\bea
	\frac{\qs^{rk}_0-\qs^{rk}_1}{2}&=& 
	\frac{1}{2}\sqrt{d\ql_+}\sqrt{d\ql_- + 2(1-\qb) -\frac{d^2}{2}(\ql_++\ql_-)}\Big(\ket A\bra C+\ket C\bra A\Big)\nn\\
	&& - \frac{1}{2}\sqrt{d\ql_-}\sqrt{d\ql_++2\qb-\frac{d^2}{2}(\ql_++\ql_-)}\Big(\ket B\bra D+\ket D\bra B\Big).
\label{sigdiff2}
\eea
\end{corollary}
\underline{\it Proof:}
Using Theorem \ref{th:sigmafull}, we see everything except the AC and BD crossterms cancel from (\ref{sigmadiag}). 
\hfill$\square$

\section{Statistical distance; proof of Theorem \ref{theorem:mainstatdist}}
\label{sec:statdist}

Now that we have described Eve's most general allowed state, and how it is connected to Alice's secret bit $s'$,
it is time to prove Theorem \ref{theorem:mainstatdist}.

Let $r_i$ be the `$r$'-value in round $i$
and similarly $k_i$, $s_i'$.
We use the notation ${\bf r}=(r_1,\ldots,r_n)$, ${\bf k}=(k_1,\ldots,k_n)$.
Let $x=(s_1',\ldots,s_n')$.
Let $z\in\bits^\ell$ be the QKD key obtained by applying privacy amplification to $x$, i.e.
$z={\tt Ext}(x,u)$, where ${\tt Ext}$ is a universal hash function (UHF) and $u\in\cU$ is public randomness.
At given $({\bf r},{\bf k})$
the quantum-classical state describing the whole system is
\bea
	\qr({\bf r},{\bf k}) &=& \EE_{z\in\bits^\ell}\EE_u\ket z\bra z \otimes\ket u\bra u\otimes\qo(z,u,{\bf r},{\bf k})
	\\ 
	\qo(z,u,{\bf r},{\bf k})&\isdef & \EE_{x\in\bits^n: {\tt Ext}(x,u)=z}
	\bigotimes_{i=1}^n \qs^{r_ik_i}_{x_i}
	\\ &=&
	2^{\ell-n}\sum_{x\in\bits^n}\qd_{z,{\tt Ext}(x,u)} \bigotimes_{i=1}^n \qs^{r_ik_i}_{x_i}.
\eea
We take the $z$-averaged of $\qo$,
\be
	\qo_{\rm av}({\bf r},{\bf k}) \isdef \EE_z \qo(z,u,{\bf r},{\bf k}) 
	= \bigotimes_{i=1}^n \frac{\qs^{r_ik_i}_0+\qs^{r_ik_i}_1}2.
\ee
Note that $\qo_{\rm av}$ does not depend on~$u$.
Furthermore we define the `ideal' decoupled state as
\be
	\qr_{\rm id}({\bf r},{\bf k})\isdef 
	\EE_{zu}\ket z\bra z\otimes \ket u\bra u\otimes \qo_{\rm av}({\bf r},{\bf k})
\ee
and we introduce the notation
$\qD(z,u,{\bf r},{\bf k})=\qo(z,u,{\bf r},{\bf k})-\qo_{\rm av}({\bf r},{\bf k})$.


We look at the security of $Z$ given ${\bf r},{\bf k},U$ and $\qo(Z,U,{\bf r},{\bf k})$.
We follow definition (\ref{defDXgivenrhoX})
and write $Z$'s distance from uniformity as
\bea
	\cD\big(Z|U,{\bf r},{\bf k},\qo(Z,U,{\bf r},{\bf k})\big)
	&=& D\big(\qr({\bf r},{\bf k}),\; \qr_{\rm id}({\bf r},{\bf k}) \big)
	\nn \\ &=&
	\frac12\big\| \EE_{zu}\ket z\bra z\otimes \ket u\bra u\otimes \qD(z,u,{\bf r},{\bf k})
	\big\|_1
\eea

\begin{lemma}
\label{lemma:blockstructure}
It holds that
\be
	\| \qr({\bf r},{\bf k})-\qr_{\rm id}({\bf r},{\bf k}) \|_1
	=\EE_{zu}\| \qD(z,u,{\bf r},{\bf k}) \|_1.
\ee
\end{lemma}
\underline{\it Proof:}
This follows from the block structure of $\qr-\qr_{\rm id}$.
The list of eigenvalues of $\qr-\qr_{\rm id}$ 
is obtained by combining 
the individual eigenvalue lists of the
$\qD(z,u,{\bf r},{\bf k})$
for all combinations $(z,u)$. 
\hfill$\square$

\begin{lemma}
\label{lemma:JensenDelta}
It holds that
\be
	\EE_{zu}\| \qD(z,u,{\bf r},{\bf k}) \|_1 \leq \tr\sqrt{\EE_{zu}\qD^2(z,u,{\bf r},{\bf k})}.
\ee
\end{lemma}

\underline{\it Proof:}
$\EE_{zu}\| \qD(z,u,{\bf r},{\bf k}) \|_1=\EE_{zu}\tr\sqrt{\qD^2(z,u,{\bf r},{\bf k})}$
$=\tr\EE_{zu}\sqrt{\qD^2(z,u,{\bf r},{\bf k})}$.
We apply Jensen's inequality.
\hfill$\square$

\begin{lemma}
\label{lemma:EDeltasq}
It holds that
\be
	\EE_{zu}\qD^2(z,u,{\bf r},{\bf k}) =
	\frac{2^\ell-1}{2^n}\bigotimes_{i=1}^n \frac{(\qs^{r_i k_i}_0)^2+(\qs^{r_i k_i}_1)^2}2.
\label{EDeltasq}
\ee
\end{lemma}
\underline{\it Proof:}
From the definition of $\qo$ and $\qo_{\rm av}$ we get
\bea
	\EE_{zu}\qD^2(z,u,{\bf r},{\bf k}) &=&
	\frac{2^{2\ell}}{2^{2n}}\sum_{xy}\EE_{zu}\qd_{z,{\tt Ext}(x,u)}\qd_{z,{\tt Ext}(y,u)}
	\bigotimes_{i=1}^n \qs^{r_i k_i}_{x_i}\qs^{r_i k_i}_{y_i}
	+\qo_{\rm av}^2
	\nn\\ && 
	-\qo_{\rm av}\frac{2^\ell}{2^n}\sum_x\EE_{zu}\qd_{z,{\tt Ext}(x,u)}\bigotimes_{i=1}^n\qs^{r_i k_i}_{x_i}
	\nn\\ &&
	-\Big(\frac{2^\ell}{2^n}\sum_x\EE_{zu}\qd_{z,{\tt Ext}(x,u)}\bigotimes_{i=1}^n\qs^{r_i k_i}_{x_i}\Big)\qo_{\rm av}.
\eea
We split the $\sum_{xy}$ sum into a sum with $y=x$ and a sum with $y\neq x$.
Then we use $\sum_z \qd_{z,{\tt Ext}(x,u)}=1$
and $\sum_z \EE_u\qd_{z,{\tt Ext}(x,u)}\qd_{z,{\tt Ext}(y,u)}=2^{-\ell}$ for $y\neq x$.
The latter is the defining property of UHFs.
Then we rewrite $\sum_{xy:\,y\neq x}$ as $\sum_{xy}-\sum_{xy}\qd_{xy}$.
Finally, after applying $2^{-n}\sum_x\bigotimes_i \qs^{r_i k_i}_{x_i}=\qo_{\rm av}$,
most of the terms cancel and (\ref{EDeltasq}) is what remains.
\hfill$\square$

\begin{lemma}
\label{lemma:avsigmasq}
It holds that
\bea
	\frac{(\qs^{r k}_0)^2+(\qs^{r k}_1)^2}2 &=&
	\sum_{{t\in\{0,\ldots,d-1\}}\atop{t\neq k,t\neq \ell}} \xi_0^2
	\Big(  \frac{\ket{w_{tk}}\bra{w_{tk}}}  {\inprod{w_{tk}}{w_{tk}}}
	+ \frac{\ket{w_{t\ell}}\bra{w_{t\ell}}} {\inprod{w_{t\ell}}{w_{t\ell}}}  \Big)
	+\xi_1(\xi_1-\fr d2\ql_-) \ket B\bra B
	\nn\\ &&
	+\xi_1\fr d2 \ql_-\ket D\bra D +\xi_2(\xi_2-\fr d2\ql_+)\ket A\bra A
	+\xi_2\fr d2 \ql_+ \ket C\bra C.
	\nn
\eea
\end{lemma}
\underline{\it Proof:}
Follows directly from Theorem~\ref{th:sigmafull}.
\hfill$\square$

\begin{lemma}
\label{lemma:boundstatlabda}
The statistical distance between the real and ideal state can be bounded as
\bea
	&& \| \qr({\bf r},{\bf k})-\qr_{\rm id}({\bf r},{\bf k}) \|_1 < \sqrt{2^{\ell-n}}T^n
	\\
	&& T \isdef 2(d-2)\xi_0+\sqrt{\xi_2(\xi_2-\fr d2\ql_+)}+\sqrt{\xi_2\fr d2\ql_+}
	+\sqrt{\xi_1(\xi_1-\fr d2\ql_-)}+\sqrt{\xi_1\fr d2\ql_-}.
	\quad\quad
\label{defT}
\eea
\end{lemma}
\underline{\it Proof:}
Substitution of Lemma~\ref{lemma:EDeltasq} into Lemma~\ref{lemma:JensenDelta} into
Lemma~\ref{lemma:blockstructure} gives
$\| \qr({\bf r},{\bf k})-\qr_{\rm id}({\bf r},{\bf k}) \|_1$
$\leq\sqrt{\frac{2^\ell-1}{2^n}}\prod_{i=1}^n \tr\sqrt{\frac{(\qs^{r_i k_i}_0)^2+(\qs^{r_i k_i}_1)^2}2}$.
The trace does not depend on the actual value of $r_i$ and $k_i$.
We define $T=\tr \sqrt{(\qs^{r k}_0)^2+(\qs^{r k}_1)^2}/\sqrt2$ for arbitrary $r,k$. 
From Lemma~\ref{lemma:avsigmasq} we obtain~(\ref{defT}).
Finally we use $2^\ell-1<2^\ell$.
\hfill$\square$

\begin{corollary}
\label{corol:distrate}
Let $\qe$ be a small constant.
The distance $\| \qr({\bf r},{\bf k})-\qr_{\rm id}({\bf r},{\bf k}) \|_1$
can be made equal to $\qe$ by setting
$\ell/n=1-2\log T-\frac2n\log\frac1\qe$.
\end{corollary}

{\it Remark.}
Corollary~\ref{corol:distrate} provides a tighter bound on the QKD rate than
similar statements based on R\'{e}nyi-2 entropy.
We are able to compute the square root in $\tr\sqrt{\qs_0^2+\qs_1^2}$, whereas
in R\'{e}nyi-2 entropy Jensen's inequality is used to bound the trace as
$\sqrt{\mbox{dimension}}\sqrt{\tr\qs_0^2+\tr\qs_1^2}$.\\

Since Eve is still free to choose the parameters $\mu$ and $V$ 
(or, equivalently, $\ql_+$ and $\ql_-$)
she can choose them such that $\| \qr({\bf r},{\bf k})-\qr_{\rm id}({\bf r},{\bf k}) \|_1$ is maximized.

\begin{theorem}
\label{th:resultdelta}
Eve's choice that maximises $\| \qr({\bf r},{\bf k})-\qr_{\rm id}({\bf r},{\bf k}) \|_1$ is given by
\bea
	\qb\leq \qb_* &: \quad& 
	T = 2\qb+\sqrt{1-2\qb}\Big[\sqrt{1-2\qb\frac{d-1}{d-2}}+\frac{\sqrt{2\qb}}{\sqrt{d-2}}\Big]
\label{Tpunt}
	\\
	&& \mbox{at } \ql_-=0, \quad \ql_+=\frac{4\qb}{d(d-2)}
\label{driehoekpunt}
	\\
	\qb\geq\qb_* &:\quad & 
	T=2\qb_*+\sqrt{1-2\qb_*}\Big[\sqrt{1-2\qb_*\frac{d-1}{d-2}}+\frac{\sqrt{2\qb_*}}{\sqrt{d-2}}\Big]
\label{Tsat}
	\\
	&& \mbox{at } \ql_-=\frac{4\qb_*(\qb-\qb_*)}{d(d-2)(1-2\qb_*)},\;\;\; 
	\ql_+= \frac{4\qb_*(1-\qb-\qb_*)}{d(d-2)(1-2\qb_*)}.
\label{Lbetalarge}
\eea
Here $\qb_*$ is a saturation value that depends on $d$ as follows,
\be
	\qb_* = \frac{x_d/2}{1+x_d},
\label{bstard}
\ee
where $x_d$ is the solution on $(0,1)$ of the equation 
\be
	(1-\frac{x}{d-2})^{\frac12}+ \frac{d-1}{d-2}(1-\frac{x}{d-2})^{-\frac12}
	+\frac1{\sqrt{d-2}}(\sqrt x-\frac1{\sqrt x})-2=0.
\label{bstarcondition}
\ee
\end{theorem}
\underline{\it Proof:}
We start from (\ref{defT}). At $\qb=\fr12$ the expression for $T$ is symmetric in $\ql_+$ and $\ql_-$.
Hence the overall maximum achievable at any $\qb$
lies at $\ql_+=\ql_-=\frac{q}{d(d-2)}$ for some as yet unknown~$q$. We have 
\be
	T^{\qb=\fr12}_{\rm max}= \qz(q,d) \isdef q +\sqrt{1-q}\Big(\sqrt{1-\frac{d-1}{d-2}q}+\frac{\sqrt q}{\sqrt{d-2}}\Big).
\ee
On the other hand, we note that substitution of (\ref{driehoekpunt}) into (\ref{defT})
yields (\ref{Tpunt}), which is precisely of the form $\qz(q,d)$ if we identify $2\qb\equiv q$.
Hence, at some $\qb<\fr12$ it is already possible to achieve $T=T^{\qb=1/2}_{\rm max}$,
i.e. we have saturation.
We note that substitution of (\ref{Lbetalarge}) into (\ref{defT}) yields (\ref{Tsat}).
The saturation value $\qb_*$ is found by solving
$\prt \qz(2\qb,d)/\prt \qb=0$; after some simplification, this equation can be rewritten as (\ref{bstarcondition})
by setting $x=2\qb/(1-2\qb)$.\footnote{
After some rewriting it can be seen that (\ref{bstarcondition})
is equivalent to a complicated 6th order polynomial equation.
We have not yet been able to prove that the solution on $(0,1)$ is unique.
Our numerical solutions however indicate that this is the case.
}
\hfill$\square$\\

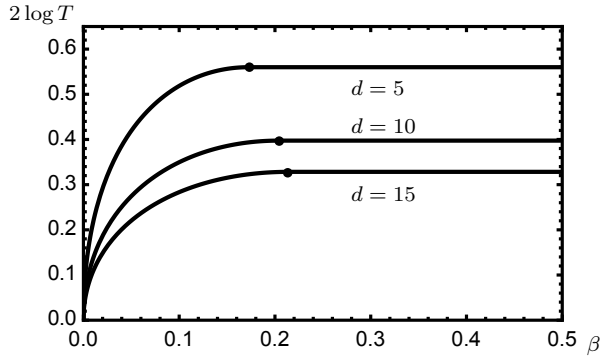
\begin{figure}[h]
\begin{center}
\begin{picture}(200,110)
\setlength{\unitlength}{1mm}
\put(0,-8){\includegraphics[width=70mm]{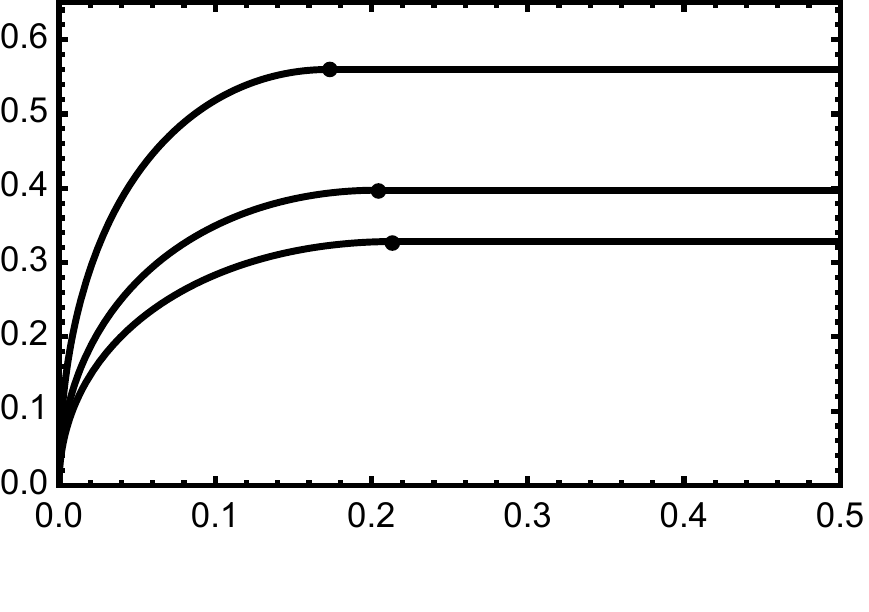}}
\put(71,-2){$\qb$}
\put(-5,42){$2 \log T$}
\put(40,32){$d=5$}
\put(40,27){$d=10$}
\put(40,18){$d=15$}
\end{picture}
\end{center}
\caption{\it
Upper bound on the information leakage as a function of the bit error rate for $d=5$, $d=10$ and $d=15$ (Theorem \ref{theorem:mainstatdist}). 
A dot indicates the saturation point $\qb_{*}$.}
\label{fig:statdist}
\end{figure}

The upper bound on the amount of information that Eve has about $S'$ is
$2 \log T$. This is a concave function of $\qb$ (see Fig.\,\ref{fig:statdist}).
Hence there is no advantage for Eve to cause different error rates in different rounds.
For Eve it is optimal to cause error rate $\qb$ in every round.

{\bf This concludes the proof of Theorem~\ref{theorem:mainstatdist}.}\\

The optimal $\ql_+$,$\ql_-$ are plotted in Fig.\,\ref{fig:opt} (Section \ref{sec:collective}).

\begin{lemma}
\label{lemma:bstarasymp}
The large-$d$ asymptotics of the saturation value $\qb_*$ is given by
\be
	\qb_* = \frac14-\frac1{8\sqrt{d-2}}-\cO(\frac1{(d-2)^{3/2}}),
\label{asympbstar}
\ee
which yields
\bea
	T &=& 1+ \frac1{2\sqrt{d-2}}-\cO(\frac1{d-2})
\label{asympT}
	\\
	\| \qr({\bf r},{\bf k})-\qr_{\rm id}({\bf r},{\bf k}) \|_1 & \leq & 2^{-\fr12n[1-\frac1{\sqrt{d-2}\ln 2}+\cO(\frac1{d-2})-\frac\ell n]}.
\label{asympnorm}
\eea
\end{lemma}
\underline{\it Proof:}
We set $x_d=1-1/\sqrt{d-2}+a/(d-2)$, where $a$ is supposedly of order~1,
and substitute this into (\ref{bstarcondition}).
This yields $a=\fr12+\cO(1/\sqrt{d-2})$, which is indeed of order~1.
Substitution of $x_d$ into (\ref{bstard}) gives (\ref{asympbstar}), and 
substitution of $\qb_*$ into (\ref{Tsat}) gives (\ref{asympT}).
Finally, substitution of (\ref{asympT}) into Lemma~\ref{lemma:boundstatlabda}
yields (\ref{asympnorm}).
\hfill$\square$


%
%
%
%

\section{Von Neumann entropy}
\label{sec:vonNeumann}

Here we prove Theorem~\ref{theorem:mainvonNeumann}.
%
%
%
%
Using smooth R\'{e}nyi entropies it was shown in \cite{RGK2005} that, in the large $n$ limit, 
the von Neumann leakage per qubit is the relevant quantity for determining the required amount of PA.\footnote{
By applying Jensen's inequality once more to lemma \ref{lemma:JensenDelta}, 
we can move the trace into the square root and get an expression which is equivalent to lemma 4.4 in \cite{RennerKonig2005}. 
After this point the proof structure from \cite{RennerKonig2005} can be followed. 
Thus the Von Neumann leakage is also an asymptotic case of our statistical distance result Theorem~\ref{theorem:mainstatdist}.
} 
We denote the leakage from Alice to Eve, in terms of von Neumann entropy, $I_{\rm AE}$. It is given by
\bea
	I_{\rm AE}&=&
	S(\qs^{RK}_{S'}|RK)- S(\qs^{RK}_{S'}|RKS')
	\nn\\ &=&
	\EE_{rk}[ S(\qs^{rk}_{S'})- S(\qs^{rk}_{S'}|S') ]
	\nn\\ &=&
	\EE_{rk}\left[ S(\frac{\qs^{rk}_0+\qs^{rk}_1}2)-\frac{S(\qs^{rk}_0)+S(\qs^{rk}_1)}2 \right]
	\nn\\ &=&
	S\Big(\frac{\qs^{rk}_0+\qs^{rk}_1}2\Big)-\frac{S(\qs^{rk}_0)+S(\qs^{rk}_1)}2 
	\quad r,k \mbox{ arbitrary}.
\label{Neumannform}
\eea
In the last line we used that the eigenvalues of $\qs^{rk}_{s'}$ 
and $\qs^{rk}_0+\qs^{rk}_1$
do not actually depend on $r$ and $k$. Again $\ql_+$ and $\ql_-$ can be optimized to Eve's advantage.

\begin{theorem}
\label{th:vonNeumann}
Eve's choice that maximizes the von Neumann leakage is given by
\bea
	\qb\leq \qb_0 &:\quad& 
	I_{\rm AE}= (1-2\qb)h(\frac1{d-2}\cdot\frac{2\qb}{1-2\qb})
\label{IAEbetasmall}
	\\ &  &  {\rm at}\;\; \ql_-=0,\;\;\; \ql_+=\frac{4\qb}{d(d-2)}
\label{cornerpoint}
	\\
	\qb\geq \qb_0 &: \quad& I_{\rm AE}=(1-2\qb_0)h(\frac1{d-2}\cdot\frac{2\qb_0}{1-2\qb_0})
\label{IAEbetalarge}
	\\ &   &{\rm at}\;\; \ql_-=\frac{4\qb_0(\qb-\qb_0)}{d(d-2)(1-2\qb_0)},\;\;\; 
	\ql_+= \frac{4\qb_0(1-\qb-\qb_0)}{d(d-2)(1-2\qb_0)}.
\label{lambdasol}
\eea
Here $\qb_0$ is a saturation value that depends on $d$ as follows,
\be
	\qb_0=\frac12\Big[1+\frac1{(d-2)(1-y_d)}\Big]^{-1}
\label{beta0}
\ee
where $y_d$ is the unique positive root of the polynomial $y^{d-1}+y-1$. 
\end{theorem}
\underline{\it Proof:} 
We start from (\ref{Neumannform}). We note that the eigenvalue set of
$(\qs^{rk}_0+\qs^{rk}_1)/2$ largely coincides with that of 
$\qs^{rk}_0$ and $\qs^{rk}_1$ (Theorem~\ref{th:sigmafull} and Corollary~\ref{corol:sigmaav}). 
What remains of (\ref{Neumannform}) comes entirely from the $\ket A,\ket B,\ket C,\ket D$ subspace,
\bea
	I_{\rm AE}&=&
	\xi_1\log \xi_1 +\xi_2\log \xi_2 -(\xi_2-\fr d2\ql_+)\log(\xi_2-\fr d2\ql_+)-\fr d2\ql_+\log(\fr d2\ql_+)
	\nn\\ &&
	-(\xi_1-\fr d2\ql_-)\log(\xi_1-\fr d2\ql_-)-\fr d2\ql_-\log(\fr d2\ql_-)
	\nn \\ &=&
	\xi_1 h(\frac d2\cdot\frac{\ql_-}{\xi_1}) + \xi_2 h(\frac d2\cdot\frac{\ql_+}{\xi_2}).
\label{h1h2form}
\eea
We note that (\ref{h1h2form}) is invariant under the transformation
$(\qb\to1-\qb; \ql_+\leftrightarrow \ql_-)$.
At $\qb=1/2$ we must hence have $\ql_+=\ql_-=\ql$.  
\be
	I_{\rm AE}^{\qb=\fr12}=g(d,\ql)
	\isdef[1-d(d-2)\ql]\cdot h\Big(\frac{d\ql}{1-d(d-2)\ql}\Big).
\label{gdl}
\ee
At $\qb=\fr12$, the largest leakage that Eve can cause is $\max_\ql g(d,\ql)$
$=g(d,\ql_*)$.\footnote{
$\frac{\prt^2 g(d,\ql)}{\prt\ql^2}=-\frac d\ql[1-d(d-2)\ql]^{-1}[1-d(d-1)\ql]^{-1}$,
hence $g$ 
is a concave function of $\ql$ on the interval $\ql\in[0,\frac{1}{d(d-1)}]$,
which interval coincides with the region allowed by the constraints on $\mu,V$.
The function $g$ has a single maximum at some point $\ql_*$.
}
Next we note that substitution of (\ref{lambdasol}) into (\ref{h1h2form}) yields
(\ref{IAEbetalarge});
this has the same form as $g(d,\ql)$
(\ref{gdl}) if we make the identification
$\ql d(d-2)=2\qb_0$.
Moreover, by setting $\qb_0=\fr12 \ql_* d(d-2)$, Eve achieves the overall maximum leakage
$g(d,\ql_*)$ already at a value of $\qb$ smaller than~$\fr12$.
Since the maximum leakage cannot decrease with $\qb$, this implies that the 
maximum leakage saturates at $\qb=\qb_0$ and stays constant at 
$I_{\rm AE}^{\rm max}(\qb)=g(d,\ql_*)$ on the interval $\qb\in[\qb_0,\fr12]$.
The value $g(d,\ql_*)$ precisely equals (\ref{IAEbetalarge}).
Next we determine the value of $\qb_0$.
Demanding $\prt g(d,\ql)/\prt \ql=0$ at $\ql=\ql_*$ yields 
\be
	\log \frac{[1-d(d-1)\ql_*]^{d-1}}{[1-d(d-2)\ql_*]^{d-2}\ql_* d}=0.
\ee
This is equivalent to the polynomial equation $y^{d-1}+y-1=0$
with $y\in[0,1]$
if we make the identification $y=1-\frac{\ql_* d}{1-\ql_* d(d-2)}=\frac{1-\ql_* d(d-1)}{1-\ql_* d(d-2)}$.
(It is readily seen that $\ql_*\in[0,\fr1{d(d-1)}]$ implies $y\in[0,1]$.)
This precisely matches (\ref{beta0}), because of the optimal choice $\qb_0=\fr12 \ql_* d(d-2)$.
By Descartes' rule of signs, the function $y^{d-1}+y-1$ has exactly one positive root.

When $\qb$ is decreased below $\qb_0$, the location $(\ql_-,\ql_+)$ of the maximum of 
the stationary point of $I_{\rm AE}$ leaves the `allowed' triangular region;
this happens at a corner of the triangle, $\ql_-=0$, $\ql_+=\frac{4\qb}{d(d-2)}$. 
For $\qb<\qb_0$ this corner yields the highest achievable leakage.
Substitution of (\ref{cornerpoint}) into (\ref{h1h2form}) yields~(\ref{IAEbetasmall}).
\hfill$\square$\\
This concludes the proof of theorem \ref{theorem:mainvonNeumann}.

Note that the leakage $I_{\rm AE}$ is a concave function of~$\qb$.
Hence it is optimal for Eve to cause error rate $\qb$ in every round.

\vskip2mm

{\it Remark}. From $y> 0$ and (\ref{beta0}) it follows that $\qb_0<\frac12\cdot\frac{d-2}{d-1}$.

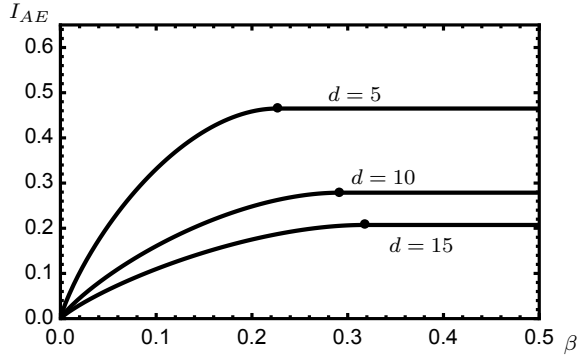
\begin{figure}[h]
\begin{center}
\begin{picture}(200,110)
\setlength{\unitlength}{1mm}
\put(0,-8){\includegraphics[width=70mm]{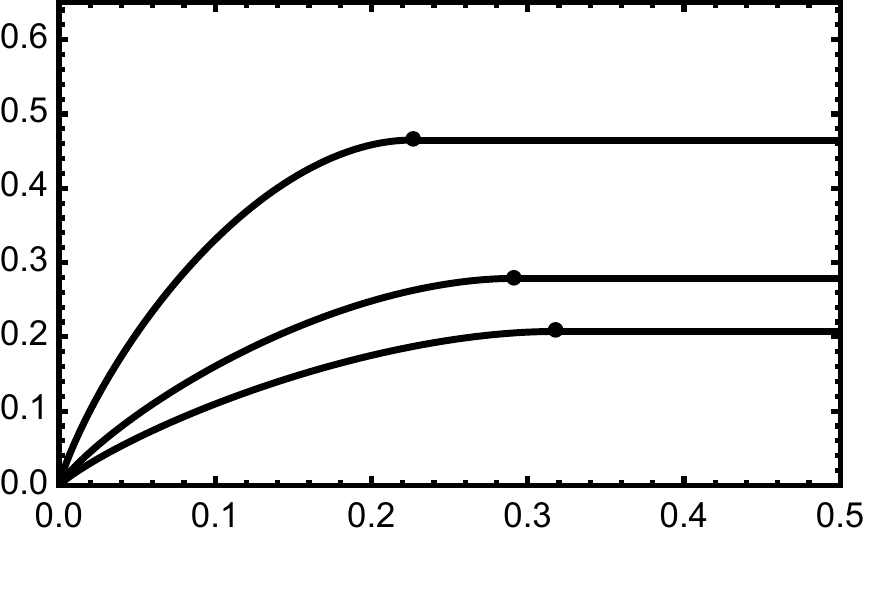}}
\put(71,-2){$\qb$}
\put(-2,42){$I_{AE}$}
\put(40,31){$d=5$}
\put(43,20){$d=10$}
\put(48,11){$d=15$}
\end{picture}
\end{center}
\caption{\it
Mutual information between Alice and Eve in terms of von Neumann entropy as a function of the bit error rate, for $d=5$, $d=10$ and $d=15$ (Theorem \ref{theorem:mainvonNeumann}). 
A dot indicates the saturation point $\qb_{0}$.}
\label{fig:VNbeta}
\end{figure}

Fig.\,\ref{fig:VNbeta} shows 
the von Neumann mutual information for three values of~$d$. 
The optimal $\ql_+$,$\ql_-$ are plotted in Fig.\,\ref{fig:opt} (Section~\ref{sec:collective}).

\begin{lemma}
The large-$d$ asymptotics of the $I_{\rm AE}$ is given by
\bea
	\qb\leq\qb_0 &:&
	I_{\rm AE}=\frac{2\qb}{d-2}\log\frac{(d-2)(1-2\qb)e}{2\qb}+\cO(d^{-2})
	\\
	\qb\geq\qb_0 &:& I_{\rm AE}=\frac{\log d}d +\cO(\frac{\log\log d}d).
\label{asympIAE}
\eea
\end{lemma}
\underline{\it Proof:} 
The result for $\qb<\qb_0$ follows by doing a series expansion of (\ref{IAEbetasmall})
in the small parameter $1/(d-2)$.
For $\qb>\qb_0$ we study the equation $y^{d-1}=1-y$.
Let us try a solution of the form $y=1-\frac{\ln[(d-1)/\qa]}{d-1}$ for some unknown $\qa$.
This yields $\qa\cdot\{(1-\frac{\ln[(d-1)/\qa]}{d-1})^{d-1}\frac{d-1}\qa\}=\ln\frac{d-1}\qa$.
Using the fact that $\lim_{n\to\infty}(1-x/n)^n=e^{-x}$ we see
 that the expression $\{\cdots\}$
is close to~1 if it holds that $\ln\frac{d-1}\qa\ll d-1$, and that the equation is then satisfied by $\qa=\cO(\ln d)$,
which is indeed consistent with $\ln\frac{d-1}\qa\ll d-1$. 
Substituting $\qa=\cO(\ln d)$ into the expression for $y$ and then into (\ref{beta0}) gives 
$1-2\qb_0=\frac1{\ln d}+\cO(\frac{\ln\ln d}{[\ln d]^2})$.
Substituting this result for $1-2\qb_0$ into (\ref{IAEbetalarge}) finally yields (\ref{asympIAE}).
\hfill$\square$

\section{Collective attacks}
\label{sec:collective}

By way of supplementary information we present a number of results about
collective attacks. These are attacks on individual qudits, 
i.e. Eve performs the same measurement on every individual ancilla that she holds.
First, this teaches us which kind of measurement is informative for Eve.
Second, it quantifies the gap between what is provable for general attacks and
what is provable for more restricted attacks.
We compute leakage in terms of min-entropy loss and in terms of
accessible (Shannon) information.
Since min-entropy is a very conservative measure we will see that the min-entropy loss
exceeds the leakage found in Theorems~\ref{theorem:mainstatdist} and \ref{theorem:mainvonNeumann}.
The main interest is in Eve's measurement itself.
The accessible information is the relevant quantity when Eve's quantum memory is short-lived,
forcing her to perform a measurement on her ancillas before she has observed Alice and Bob's
usage of the QKD key.
As expected, the accessible information will turn out to be smaller than
the leakage of Theorems~\ref{theorem:mainstatdist} and \ref{theorem:mainvonNeumann}.

\subsection{Min-entropy}
\label{sec:Hmin}


Eve's ability to distinguish between the cases $s'=0$ and $s'=1$ depends on the distance
between $\qs^{rk}_0$ and $\qs^{rk}_1$
(see Section~\ref{sec:prelimHmin}). Eq.\,(\ref{binaryHmin}) with $p_0=\fr12$, $p_1=\fr12$ tells us that the relevant quantity 
is $\|\qs^{rk}_0-\qs^{rk}_1 \|_1$.
%
For notational convenience we define the value $\qb_{\rm{sat}}$,
\be
\qb_{\rm{sat}} \isdef \frac{1}{4}\cdot \frac{d-2}{d-1}.
\ee
Again we optimize $\ql_+$ and $\ql_-$.
\begin{lemma}
\label{lem:opt}
For all $r,k$
the choice for $\ql_+$ and $\ql_-$ that maximizes the trace distance $\frac{1}{2}\left\| \qs^{rk}_0-\qs^{rk}_1 \right\|_1$ is
\bea
\label{eq:optql}
	&\ql_+=\frac{4\qb}{d(d-2)}  \hskip30mm
	\ql_- = 0 & 
	\quad\mbox{for } \qb < \qb_{\rm{sat}}
\\
	&\ql_+=\frac{4\qb_{\rm{sat}}}{d(d-2)}-\frac{2(\qb-\qb_{\rm{sat}})}{d^2} \quad\quad
	\ql_-=\frac{2(\qb-\qb_{\rm{sat}})}{d^2} &
	\quad\mbox{for } \qb \geq \qb_{\rm{sat}}.
\label{eq:optql2}
\eea
which gives
\be
\frac{1}{2}\left\| \qs^{rk}_0-\qs^{rk}_1 \right\|_1 =  
\left\{
  \begin{array}{@{}ll@{}}
      \frac{1}{\sqrt{d-1}} \frac{\sqrt{\beta}}{\qb_{\rm{sat}}}\sqrt{2\qb_{\rm{sat}}-\beta} &\mbox{for} \quad \qb < \qb_{\rm{sat}}\\ \\
      \frac{1}{\sqrt{d-1}} & \mbox{for}\quad \qb \geq \qb_{\rm{sat}}.
  \end{array}\right.
\label{resulttracedist}
\ee
\end{lemma}
\underline{\it Proof:}
From Corollary \ref{corol:sigdiff2} it is easy to see that
\bea
\label{eq:lambdasom}
\frac{1}{2} \left\| \qs^{rk}_0-\qs^{rk}_1 \right\|_1 &=&  \sqrt{d\ql_-}\sqrt{d\ql_++2\qb-\frac{d^2}{2}(\ql_++\ql_-)}\nn\\
&&+\sqrt{d\ql_+}\sqrt{d\ql_- + 2(1-\qb) -\frac{d^2}{2}(\ql_++\ql_-)}.
\eea
In Appendix~\ref{app:Hminopt} we derive the $\lambda_+$, $\ql_-$ that maximize (\ref{eq:lambdasom}) while keeping all 
eigenvalues non-negative.
\hfill$\square$\\

{\it Remark.}
The optimal choice for $\ql_+$,$\ql_-$ has the same form for all three optimizations that we have performed. 
The only difference is the saturation value. 
Although (\ref{eq:optql2}) is shown in a simplified form one can manipulate it to the same form as (\ref{Lbetalarge}) and (\ref{lambdasol}) with $\qb_{\rm sat}$ instead of $\qb_*$ or $\qb_0$.\\

\begin{figure}[h]
\begin{center}
\begin{picture}(165,160)
\setlength{\unitlength}{1mm}
\put(0,-6){\includegraphics[width=50mm]{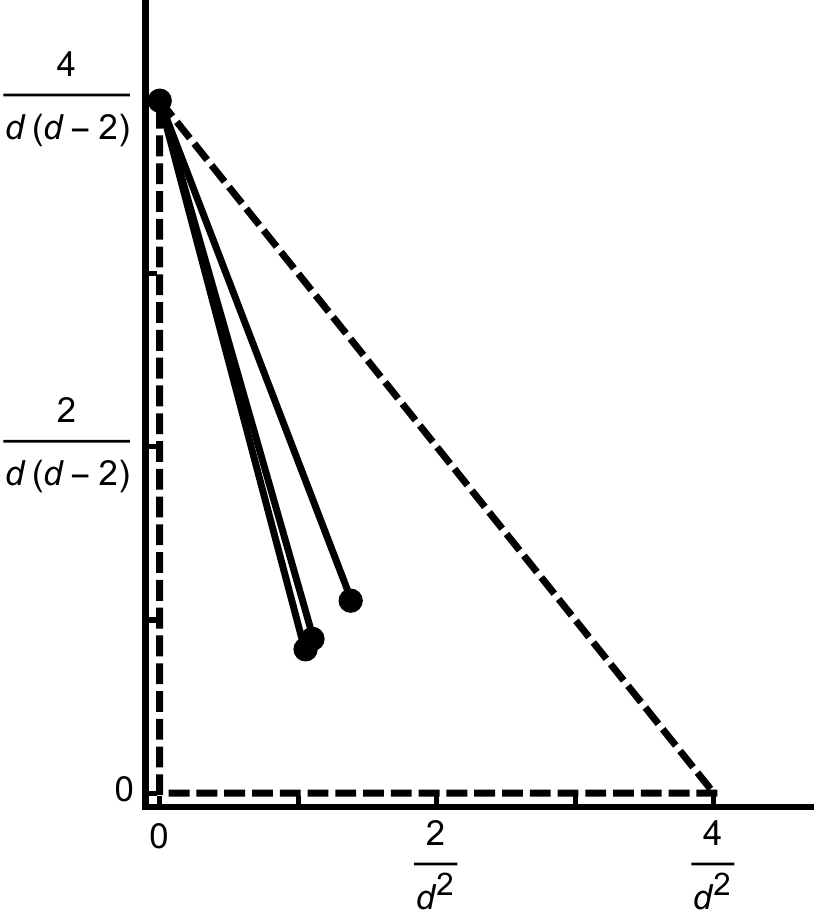}}
\put(51,0){$\ql_-/\qb$}
\put(7,52){$\ql_+/\qb$}
\end{picture}
\end{center}
\caption{\it 
Optimal choice of $\ql_+$ and $\ql_-$ at $d=10$ for statistical distance (left line), min-entropy (middle line) and von Neumann entropy (right line).
The dashed triangle represents the region for which the eigenvalues $\ql_+,\ql_-$ and $\ql_1$ are non-negative. 
The black dots indicate the optimum at $\beta=\frac{1}{2}$ (dots inside the triangle) 
and $\qb \leq \qb_*, \qb_{\rm{sat}}, \qb_0$ (upper corner of the triangle). 
Not shown in this plot is the $\ql_0 \geq 0$ constraint which cuts off the upper left corner of the triangle for  $\qb>2\qb_{\rm{sat}}$.}
\label{fig:opt}
\end{figure}

Fig.\,\ref{fig:opt} shows the optimal $\ql_+$ and $\ql_-$ together with the constraints on the $\ql$ parameters for all three optimizations. The lower dots in the figure correspond to $\qb=\frac{1}{2}$. 
For all three information measures the optimum moves towards the top corner of the triangle for decreasing $\qb$. 
For $\qb$ values below the saturation point the optimum is the top corner, with $\ql_-=0$ and $\ql_1=0$.\\ 

Knowing the optimal values for $\ql_+$ and $\ql_-$, we compute the min-entropy leakage.

\begin{theorem}
\label{theorem:min}
The min-entropy of the bit $S'$ given $R,K$ and the state $\qs^{RK}_{S'}$ is
\bea
\label{resultHminlow}
	\qb <\qb_{\rm{sat}}: \quad \Hmin(S'|RK \qs^{RK}_{S'}) &=&  
	- \log \left(\frac12 + \frac{1}{2\sqrt{d-1}} \frac{\sqrt{\beta}}{\qb_{\rm{sat}}}\sqrt{2\qb_{\rm{sat}}-\beta}\right)
	\quad\quad
	\\
	\qb \geq \qb_{\rm{sat}}: \quad \Hmin(S'|RK \qs^{RK}_{S'}) &=&  
	- \log \left(\frac12 + \frac{1}{2\sqrt{d-1}} \right).
\label{resultHminsat}
\eea
\end{theorem}
\underline{\it Proof:}
Eq.\,(\ref{binaryHmin}) with $X$ uniform, $X\to S'$, $Y\to (R,K)$ becomes
\bea
\label{eq:HminRRDPS}
	\Hmin(S'|RK\qs^{RK}_{s'}) &=& 
	-\log \left(\frac12+\frac12\EE_{rk} \Big\| \frac{1}{2}\qs^{rk}_0 - \frac{1}{2} \qs^{rk}_1 \Big\|_1 \right)
	\nn\\ &=& 
	-\log \left(\frac12+ \frac{1}{4} \left\| \qs^{rk}_0 - \qs^{rk}_1 \right\|_1 \right)
\quad\quad (r,k\mbox{ arbitrary}).
\eea
In the last step we omitted the expectation over $r$ and $k$ since the trace distance does not depend on $r,k$. 
Substitution of (\ref{resulttracedist}) into (\ref{eq:HminRRDPS}) gives the end result.
\hfill$\square$

\begin{corollary}
\label{optPOVM}
Eve's optimal POVM $\cT^{rk}=(T^{rk}_0,T^{rk}_1)$ for maximising the min-entropy leakage is given by
\be
	T^{rk}_0 = \frac{1}{2} \Big(\one + \ket{A}\bra{C} +\ket{C}\bra{A} - \ket{B}\bra{D} - \ket{D}\bra{B} \Big)
	\quad ; \quad 
	T^{rk}_1=\one-T^{rk}_0.
\label{T0T1opt}
\ee
\end{corollary}
\underline{\it Proof:}
The trace distance in Lemma~\ref{lem:opt} is the sum of the positive eigenvalues of $\qs^{rk}_0-\qs^{rk}_1$. 
In the space spanned by $\ket A, \ket B, \ket C, \ket D$,
the optimal $T_0$ consists of the projection onto the space spanned by the eigenvectors corresponding to the positive eigenvalues. 
These eigenvectors are $\ket{v_1}=\frac{\ket{A}+\ket{C}}{\sqrt{2}}$ and $\ket{v_2}=\frac{\ket{D}-\ket{B}}{\sqrt{2}}$. 
The matrix that projects onto them is 
$\ket{v_1}\bra{v_1}+\ket{v_2}\bra{v_2}=\fr12\ket A\bra A+\fr12\ket B\bra B+\fr12\ket C\bra C+\fr12\ket D\bra D$
$+\ket A\bra C+\ket C\bra A-\ket B\bra D-\ket D\bra B$. 
In order to satisfy the constraint $T_0+T_1=\one$ and symmetry, half the identity matrix in the remaining 
$d^2-4$ dimensions has to be added to~$T_0$. 
We mention, without showing it, that (\ref{T0T1opt}) satisfies the test (\ref{Holevotest}).
\hfill$\square$\\

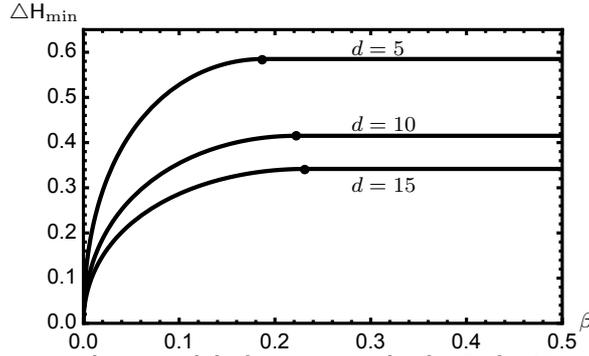
\begin{figure}[h]
\begin{center}
\begin{picture}(200,110)
\setlength{\unitlength}{1mm}
\put(0,-6){\includegraphics[width=70mm]{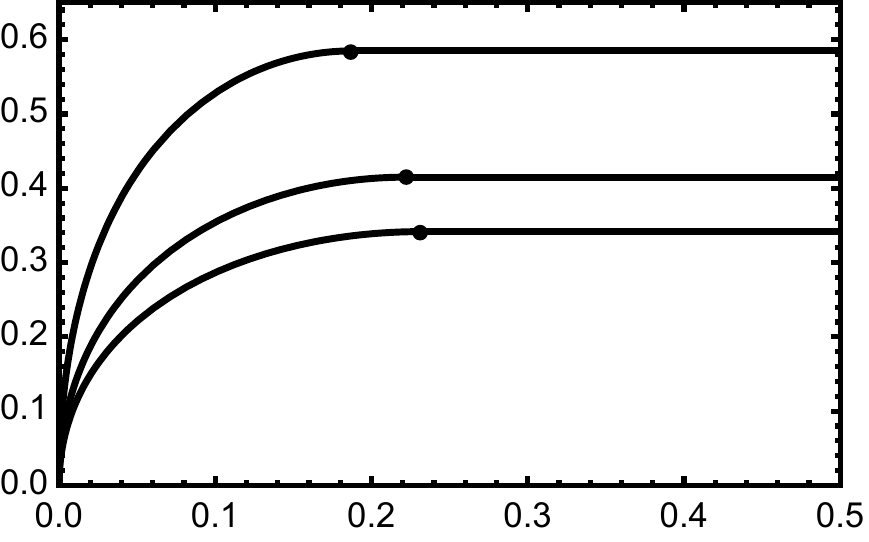}}
\put(70,-2){$\qb$}
\put(-5,39){$\tri \sH_{\rm min}$}
\put(40,34){$d=5$}
\put(40,24){$d=10$}
\put(40,16){$d=15$}
\end{picture}
\end{center}
\caption{\it
Min-entropy leakage as a function of the bit error rate for $d=5$, $d=10$ and $d=15$. 
A dot indicates the saturation point $\qb_{\rm sat}$.}
\label{fig:Hmin}
\end{figure}

As expected, the min-entropy loss decreases as the dimension of the Hilbert space grows. 
We see that the entropy loss saturates at $\qb = \qb_{\rm{sat}}$;
hence RRDPS is secure up to arbitrarily high noise levels. 
Fig.\,\ref{fig:Hmin} shows the min-entropy leakage as a function of $\qb$. 

\subsection{Accessible Shannon information}
\label{sec:Shannon}

\begin{lemma}
\label{lemma:ShannonGen}
Let $X\in\cX$ be a uniformly distributed random variable. Let $Y\in\cY$ be a random variable.
Let $\qr_{xy}$ be a quantum state coupled to the classical $x,y$.
The Shannon entropy of $X$ given a state $\qr_{XY}$ that has to be measured (for unknown $X$ and $Y$) is given by
\be
	\sH(X|\qr_{XY})=\min_{\mbox{POVM }\cM=(M_m)_{m\in\cX}} \EE_{x\in\cX}
	\sH\Big(\{ \tr M_m \EE_{y|x}\qr_{xy} \}_{m\in\cX}\Big).
\ee
\end{lemma}
\underline{\it Proof:}
We have $\sH(X|\qr_{XY})=\min_\cM \sH(X|Z)$, where $Z$ is the outcome of the POVM measurement~$\cM$.
$Z$ is a classical random variable that depends on $X$ and $Y$.
We can write $\sH(X|Z)=\sH(X)-\sH(Z)+\sH(Z|X)$.
Since $X$ is uniform, and $Z$ is an estimator for $X$, the $Z$ is uniform as well.
Thus we have $\sH(X)-\sH(Z)=0$, which yields
$\sH(X|\qr_{XY})=\min_\cM \sH(Z|X) = \min_\cM \EE_x \sH(Z|X=x)$.
The probability $\pr[z|x]$ is given by $\pr[z|x]=\EE_{y|x}\pr[z|xy]=\EE_{y|x}\tr M_z\qr_{xy}$.
\hfill$\square$

\begin{corollary}
\label{corol:HSprime}
It holds that
\bea
	\sH(S'|RK\qs^{RK}_{AS})&=&\EE_{rk}\min_{\cG^{rk}=(G^{rk}_0,G^{rk}_1)} \!\!\!  \EE_{s'}
	h(\tr G^{rk}_m \qs^{rk}_{s'}),\quad
	m\in\bits\mbox{ arbitrary}.\quad\quad
\label{eqcorol:HSprime}
\eea

\end{corollary}
\underline{\it Proof:}
Application of Lemma~\ref{lemma:ShannonGen} yields
\bea
	\sH(S'|RK\qs^{RK}_{AS}) &=& \EE_{rk}\min_{\cG^{rk}=(G^{rk}_0,G^{rk}_1)}\EE_{s'}
	H(\{\tr G^{rk}_m \EE_{as|s'}\qs^{rk}_{as}\}_{m\in\bits})
	\nn\\ &=&
	\EE_{rk}\min_{\cG^{rk}=(G^{rk}_0,G^{rk}_1)}\EE_{s'}
	H(\{\tr G^{rk}_m \qs^{rk}_{s'}\}_{m\in\bits})
\eea
where in the last step we used the definition of $\qs^{rk}_{s'}$.
Finally, the Shannon entropy of a binary variable is given by the binary entropy function $h$,
where $h(1-p)=h(p)$.
\hfill$\square$

From Corollary \ref{corol:HSprime} we see that the POVM $\cT^{rk}$ associated with the min-entropy
also optimizes the Shannon entropy: maximizing the guessing probability
$\tr G^{rk}_{s'}\qs^{rk}_{s'}$  minimizes the Shannon entropy.

\begin{theorem}
\label{theorem:Shannon}
The Shannon entropy of Alice's bit $S'$ given the state $\qs^{RK}_{AS}$, $R$ and $K$ is:
\bea
	\qb <\qb_{\rm{sat}}: \quad\quad 
	\sH(S'|RK\qs^{RK}_{AS}) &=& 
	h\left( \frac{1}{2}  + \frac{1}{2\sqrt{d-1}} \frac{\sqrt{\qb}}{\qb_{\rm{sat}}} \sqrt{2\qb_{\rm{sat}} - \qb} \right).
	\\
	\qb \geq \qb_{\rm{sat}}: \quad\quad 
	\sH(S'|RK\qs^{RK}_{AS}) &=&  h \left( \frac{1}{2}  + \frac{1}{2\sqrt{d-1}} \right).
\eea
\end{theorem}
\underline{\it Proof:}
The min-entropy result (\ref{resultHminlow},\ref{resultHminsat})
can be written as $\Hmin(S'|RK\qs^{RK}_{S'}) = -\log \tr T^{rk}_{s'} \qs^{rk}_{s'}$, 
so we already have an expression for $\tr T^{rk}_{s'} \qs^{rk}_{s'}$.
Substitution of $\cT^{rk}$ for $\cG^{rk}$
in (\ref{eqcorol:HSprime}) yields the result.
\hfill $\square$

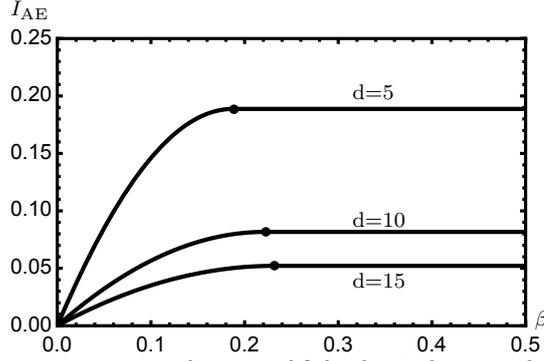
\begin{figure}[h]
\begin{center}
\begin{picture}(250,106)
\setlength{\unitlength}{1mm}
\put(0,-5){\includegraphics[width=70mm]{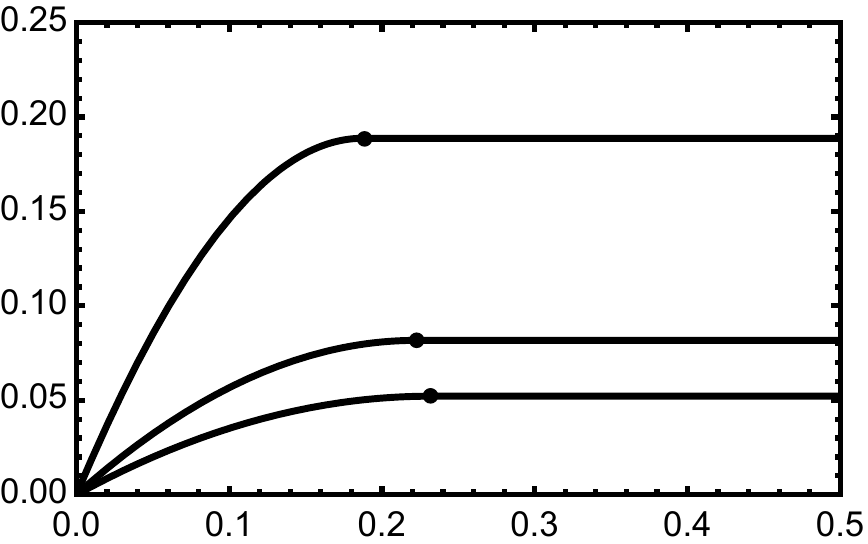}}
\put(69,-1){$\qb$}
\put(0,40){$I_{\rm AE}$}
\put(45,29){d=5}
\put(45,12){d=10}
\put(45,4){d=15}
\end{picture}
\end{center}
\caption{\it
Accessible Shannon entropy as a function of $\qb$ for $d=5$, $d=10$ and $d=15$. 
A dot indicates the saturation point $\qb_{\rm sat}$.}
\label{fig:Shannon}
\end{figure}

Since the optimal POVM for min- and Shannon entropy are the same, saturation occurs at the same point ($\qb=\qb_{\rm{sat}}$). 
Fig\,\ref{fig:Shannon} shows the Shannon entropy leakage (mutual information) 
$I_{\rm AE}=1-\sH(S'|RK\qs^{RK}_{AS})$
as a function of~$\qb$. 

\section{Discussion}
\label{sec:discussion}

\subsection{Comparison with previous analyses}
\label{sec:comparison}

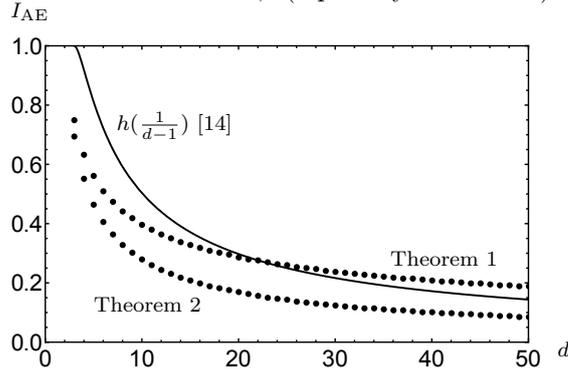
\begin{figure}[h]
\begin{center}
\begin{picture}(250,105)
\setlength{\unitlength}{1mm}
\put(0,-5){\includegraphics[width=70mm]{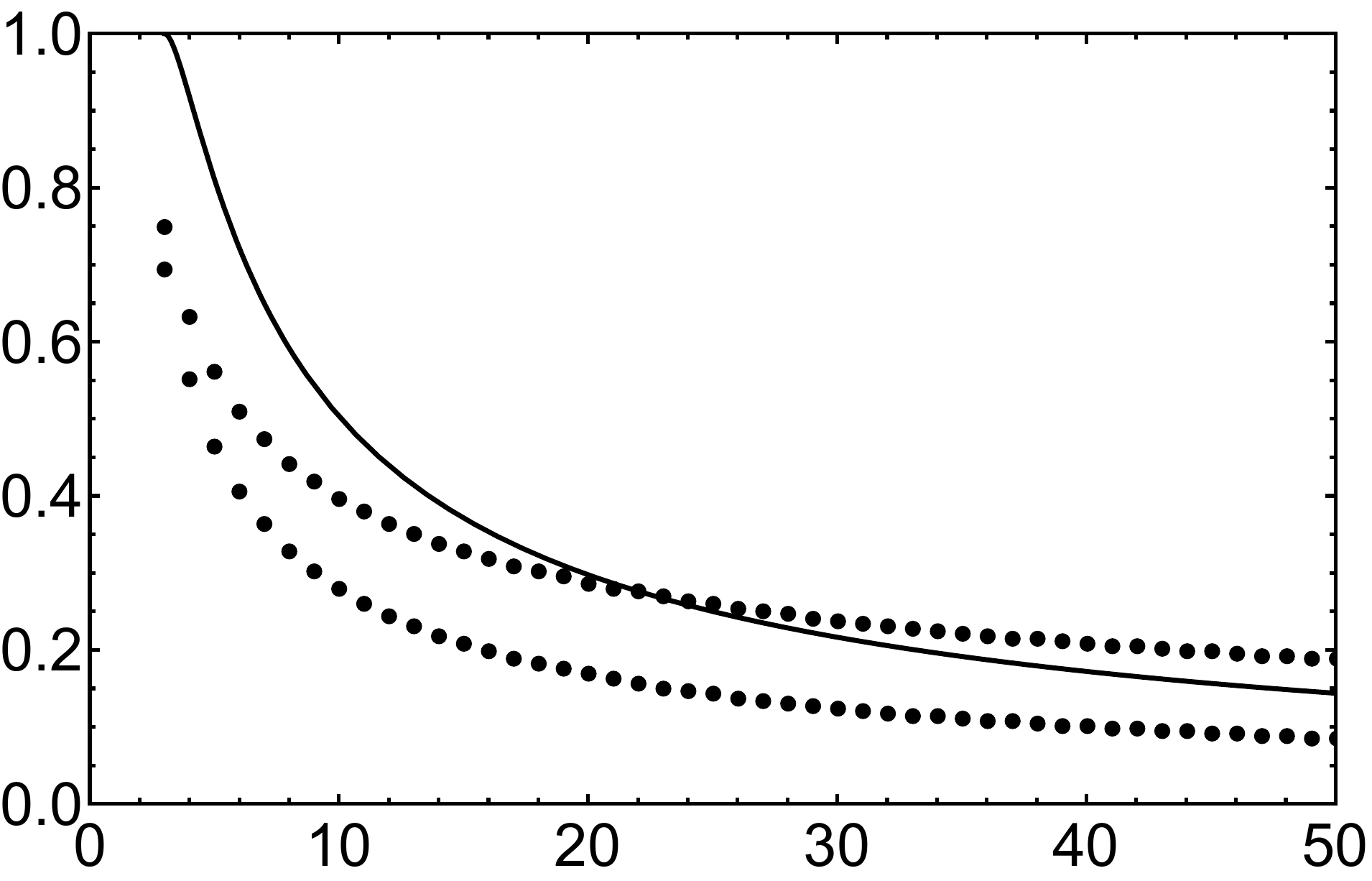}}
\put(72,-3){$d$}
\put(0,42){$I_{\rm AE}$}
\put(14,27){$h(\fr1{d-1})$ \cite{SYK2014}}
\put(50,9){${\rm Theorem}~\ref{theorem:mainstatdist}$}
\put(11,3){${\rm Theorem}~\ref{theorem:mainvonNeumann}$}
\end{picture}
\end{center}
\caption{\it
Saturated leakage as a function of $d$.
Comparison of \cite{SYK2014} and our results (Theorem~\ref{theorem:mainstatdist} and Theorem~\ref{theorem:mainvonNeumann}).}
\label{fig:compstaturated}
\end{figure}

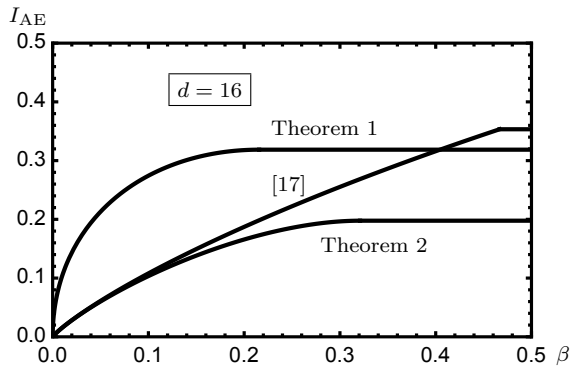
\begin{figure}[h]
\begin{center}
\begin{picture}(250,105)
\setlength{\unitlength}{1mm}
\put(0,-7){\includegraphics[width=70mm]{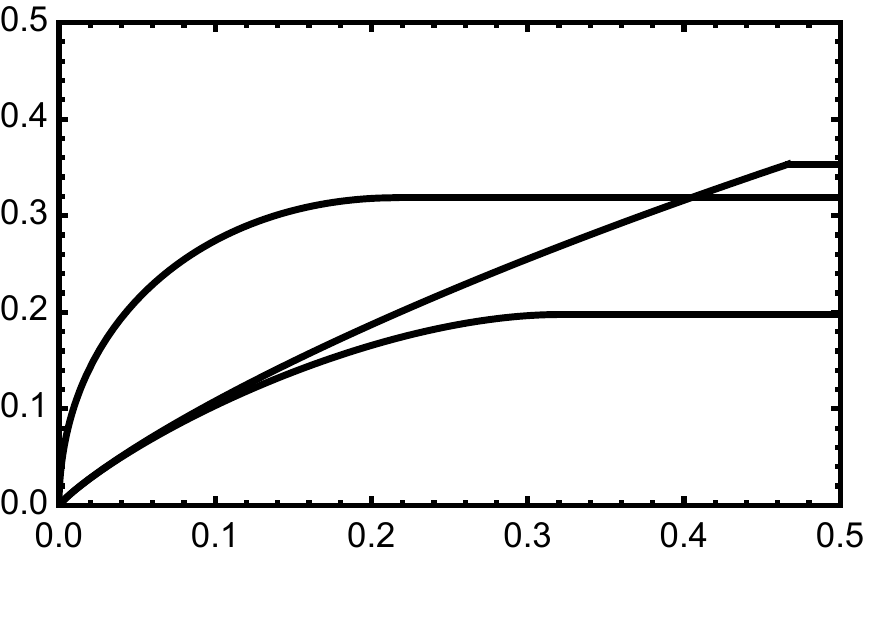}}
\put(20,35){\fbox{$d=16$}}
\put(71,0){$\qb$}
\put(-1,45){$I_{\rm AE}$}
\put(33.5,22.5){\cite{SK2017}}
\put(33.5,30){{\rm Theorem}~\ref{theorem:mainstatdist}}
\put(40,14.5){{\rm Theorem}~\ref{theorem:mainvonNeumann}}
\end{picture}
\end{center}
\caption{\it
Leakage as a function of $\qb$, for $d=16$.
Comparison of our Theorem~\ref{theorem:mainstatdist} and Theorem~\ref{theorem:mainvonNeumann}
versus
\cite{SK2017}, $h(\fr{2\qb}{d-2})$ below and 
$h(\frac{1}{d-1})$ above saturation.}
\label{fig:compbeta}
\end{figure}

Our Theorem~\ref{theorem:mainstatdist} is non-asymptotic;
we cannot compare it to previous results since the previous results are 
for the asymptotic regime.
Figs.\,\ref{fig:compstaturated} and \ref{fig:compbeta}
show our results versus
previous bounds on the leakage.
It is clear that our on Neumann result is sharper than \cite{SK2017}
for all $\qb$ and~$d$.
Interestingly, our non-asymptotic result for the saturated leakage is sharper than 
the asymptotic \cite{SK2017} for $d\leq 22$.
Note too that saturation occurs at lower $\qb$ (especially for small $d$) than reported in \cite{SK2017}.

%

\clearpage

\subsection{Remarks on the optimal attack}

The $\bar\qr^{\rm AB}$ mixed state allowed by the noise constraint has two degrees of freedom,
$\mu$ and $V$.
While this is more than the zero degrees of freedom in the case of qubit-based QKD \cite{RGK2005}, 
it is still a small number, given the dimension $d^2$ of the Hilbert space.

Eve's attack has an interesting structure.
Eve entangles her ancilla with Bob's qudit.
Bob's measurement affects Eve's state.
When Bob reveals $r,k$, Eve knows which 4-dimensional subspace is relevant.
However, the basis state $\ket k$ in Bob's qudit is coupled to $\ket{A^a_k}$ in Eve's space (see appendix \ref{app:BE}),
which is spanned by $d-1$ different basis vectors
$\ket{E^+_{(kt')}}$ (Eq.\,\ref{defA} with $\ql_1=0$, $\ql_-=0$), 
each carrying different phase information $a_k\oplus a_{t'}$.
Only one out of $d-1$ carries the information she needs, and she cannot select which one to read out.
Her problem is aggravated by the fact that the $\ket{A^a_t}$ vectors are not orthogonal (except at $\qb=\fr12$).
Note that this entanglement-based attack is far more powerful than the intercept-resend attack
studied in \cite{Sko2017}.

%
%
%

\section*{Acknowledgements}
We thank Serge Fehr for useful discussions.
We thank the anonymous reviewers of Eurocrypt 2018
for their comments about symmetrisation.
Part of this research was funded by NWO (CHIST-ERA project ID\underbar{$\;\;$}\,IOT).


\appendix

\section{Details of Eve's unitary operation}
\label{app:BE}
In Theorem \ref{th:UBE} below we show that Eve does not have to touch Alice's qudit.
Hence the attacks that we are describing here can also be carried out in the original (non-EPR)
protocol, where Eve gets access only to the qudit state sent to Bob.

\begin{theorem}
\label{th:UBE}
The operation that maps the pure EPR state to $\ket{\qJ^{\rm ABE}}$ (\ref{PsiABE})
can be represented as a unitary operation on Bob's subsystem and Eve's ancilla.
\end{theorem}
\underline{\it Proof:} 
Let Eve's ancilla have initial state $\ket{E_0}$.
The transition from the pure EPR state to (\ref{PsiABE}) can be written as the following mapping,
\be
	U\Big( \ket t_{\rm B}\otimes \ket{E_0}_{\rm E} \Big) = \ket{\qO_t},
\label{mapBE}
\ee
where $\ket{\qO_t}$ is a state in the BE system defined as
\be
	\ket{\qO_t} \isdef \sqrt{\ql_0}\ket t\ket{E_0}+\sqrt{\ql_1}\ket t\sum_{j=1}^{d-1}e^{i\frac{2\pi}d jt}\ket{E_j}
	+\sqrt{\frac{d\ql_+}2}\sum_{t':t'\neq t}\ket{t'}\ket{E^+_{(tt')}}
	+\sqrt{\frac{d\ql_-}2}\sum_{t':t'\neq t}\ket{t'}\ket{E^-_{(tt')}}\sgn(t'-t).
\label{defOmega}
\ee
The notation $(tt')$ indicates ordering of $t$ and $t'$ such that the smallest index occurs first.
It holds that $\inprod{\qO_t}{\qO_\qt}=\qd_{t\qt}$.
Eqs.\,(\ref{mapBE},\ref{defOmega}) show that the attack can be represented as an operation that does not touch Alice's subsystem.
Next we have to prove that the mapping is unitary.
The fact that $\inprod{\qO_t}{\qO_\qt}=\qd_{t\qt}$ shows that orthogonality in Bob's space is correctly preserved.
In order to demonstrate full preservation of orthogonality we have to define the action
of the operator $U$ on states of the form $\ket t_{\rm B}\otimes\ket{\qe}_{\rm E}$,
where $\ket\qe$ is one of Eve's basis vectors orthogonal to $\ket{E_0}$,
in such a way that the resulting states are mutually orthogonal and orthogonal to all $\ket{\qO_t}$,
$t\in\{0,\ldots,d-1\}$.
The dimension of the BE space is $d^3$ and allows us to make such a choice of $d(d^2-1)$ vectors.
\hfill$\square$

\begin{theorem}
\label{th:UmuaE0}
Let Alice send the state $\ket{\mu_a}$ to Bob. Let Eve apply the unitary operation $U$ (specified in the proof of Theorem~\ref{th:UBE})
to this state and her ancilla. The result can be written as
\be
	U\Big(\ket{\mu_a}\otimes\ket{E_0}\Big)=\frac1{\sqrt d}\sum_{t=0}^{d-1}(-1)^{a_t}\ket t\otimes \ket{A^a_t},
\label{UmuaE0}
\ee
\be
	\ket{A^a_t}\isdef \sqrt{\ql_0}\ket{E_0}+\sqrt{\ql_1}\sum_{j=1}^{d-1}e^{i\frac{2\pi}d jt}\ket{E_j}
	+\sqrt{\frac d2}\sum_{t':t'\neq t} \!\!\! (-1)^{a_t+a_{t'}}
	\!\! \left[
	\sqrt{\ql_+}\ket{E^+_{(tt')}} + \sqrt{\ql_-}\sgn(t'-t)\ket{E^-_{(tt')}}
	\right]\!\!.
\label{defA}
\ee
The states $\ket{A^a_t}$ are normalised and satisfy 
$\forall_{t\qt:\qt\neq t}\;\inprod{A^a_\qt}{A^a_t}=(1-2\qb)$.
\end{theorem}
\underline{\it Proof:} 
We start from $U(\ket{\mu_a}\ket{E_0})=(1/\sqrt d)\sum_t(-1)^{a_t}\ket{\qO_t}$
and we substitute (\ref{defOmega}).
Re-labeling of summation variables yields (\ref{UmuaE0},\ref{defA}).
The norm $\inprod{A^a_t}{A^a_t}$ equals $\ql_0+(d-1)\ql_1+\frac{d(d-1)}2\ql_+
+\frac{d(d-1)}2\ql_-$, which equals 1 since this is also equal to the trace of $\tilde\qr^{\rm AB}$.
For $\qt\neq t$ the inner product $\inprod{A^a_\qt}{A^a_t}$ yields
\be
	\ql_0+\ql_1\sum_{j=1}^{d-1}e^{i\frac{2\pi}d j(t-\qt)}
	+\frac d2\sum_{t'\neq t}\sum_{\qt'\neq \qt}
	(-1)^{a_t+a_{t'}+a_\qt+a_{\qt'}}\qd_{t'\qt}\qd_{\qt' t}[\ql_++\ql_-\sgn(t'-t)\sgn(\qt'-\qt)].
\label{interimA}
\ee
We use $\sum_{j=1}^{d-1}e^{i\frac{2\pi}d j(t-\qt)}=d\qd_{\qt t}-1=-1$.
Furthermore the Kronecker deltas in (\ref{interimA}) set the phase $(-1)^{\cdots}$ to~$1$
and $\sgn(t'-t)\sgn(\qt'-\qt)=\sgn(\qt-t)\sgn(t-\qt)=-1$.
Finally we use $\ql_0-\ql_1=1-2\qb-V$ and $\ql_+-\ql_-=2V/d$.
\hfill$\square$

Theorem~\ref{th:UmuaE0} reveals an intuitive picture.
In the noiseless case ($\qb=0$) it holds that $\forall_t\; \ket{A^a_t}=\ket{E_0}$,
i.e.\,Eve does nothing, resulting in the factorised state $\ket{\mu_a}\ket{E_0}$.
In the case of extreme noise ($\qb=\fr12$) we have $\inprod{A^a_t}{A^a_\qt}=\qd_{t\qt}$,
which corresponds to a maximally entangled state between Bob and Eve.

\begin{corollary}
\label{corol:rhoBob}
The pure state (\ref{UmuaE0}) in Bob and Eve's space
gives rise to the following mixed state $\qr^{\rm B}_a$ in Bob's subsystem,
\be
	\qr^{\rm B}_a=(1-2\qb)\ket{\mu_a}\bra{\mu_a}+2\qb\frac\one d.
\ee
\end{corollary}
\underline{\it Proof:}
Follows directly from (\ref{UmuaE0}) by tracing out Eve's space and
using the inner product $\inprod{A^a_\qt}{A^a_t}=(1-2\qb)$ for $\qt\neq t$.
\hfill$\square$

From Bob's point of view, what he receives is a mixture of the $\ket{\mu_a}$ state and the fully mixed state.
The interpolation between these two is linear in~$\qb$.
Note that the parameters $\mu,V$ are not visible in~$\qr^{\rm B}_a$.

\section{Optimization for the min-entropy}
\label{app:Hminopt}

Here we prove that (\ref{eq:optql},\ref{eq:optql2}) maximizes (\ref{eq:lambdasom}).
We first show that (\ref{eq:lambdasom}) is concave
and obtain the optimum for $\qb\geq\qb_{\rm sat}$.
Then we take into account the constraints on the eigenvalues and derive the optimum for $\qb<\qb_{\rm sat}$.

{\bf Unconstrained optimization}.
For notational convenience we define
\bea
	w_1 &=& \sqrt{d\ql_+ + 2\qb -\frac{d^2}{2}(\ql_+ + \ql_-)}\\
	w_2 &=& \sqrt{d\ql_- + 2(1-\qb) -\frac{d^2}{2}(\ql_+ + \ql_-)}.
\eea
This allows us to formulate everything in terms of $\ql_+$ and $\ql_-$.
Eq.\,(\ref{eq:lambdasom}) becomes
\be
\label{eq:lambdasom2}
	\frac{1}{2} \left\| \qs^{rk}_0-\qs^{rk}_1 \right\|_1=  \sqrt{d\ql_-}w_1+\sqrt{d\ql_+}w_2.
\ee
Next we compute the derivatives,
\bea
\label{eq:ddx}
	\frac{\partial}{\partial \ql_+}  \left\| \qs^{rk}_0-\qs^{rk}_1 \right\|_1 &=& 
	-\frac{d^2}{2}\frac{\sqrt{\lambda _+}}{w_2} +\frac{w_2}{\sqrt{\lambda _+}}+ (d-\frac{d^2}{2})\frac{\sqrt{\ql_-}}{w_1}.
	\\
	\frac{\partial}{\partial \ql_-}  \left\| \qs^{rk}_0-\qs^{rk}_1 \right\|_1 &=& 
	-\frac{d^2}{2}\frac{\sqrt{\lambda _-}}{w_1} +\frac{w_1}{\sqrt{\lambda _-}}+ (d-\frac{d^2}{2}) \frac{\sqrt{\ql_+}}{w_2}.
\label{eq:ddy}
\eea
Setting both these derivatives to zero yields a stationary point of the function.
Setting
$w_1\sqrt{\ql_+}\frac{\partial}{\partial \ql_+}\left\| \qs^{rk}_0-\qs^{rk}_1 \right\|_1$ 
$-w_2\sqrt{\ql_-}\frac{\partial}{\partial \ql_-}  \left\| \qs^{rk}_0-\qs^{rk}_1 \right\|_1$ 
to zero gives 
$\ql_+w_1^2-\ql_-w_2^2 =0$,
which describes a hyperbola
\be
	(\fr12 d^2-d)(\ql_-^2-\ql_+^2) +2\qb\ql_+-2(1-\qb)\ql_-  =0.
\ee
Next, the equations  
$\sqrt{\ql_+} w_1 w_2 \frac{\partial}{\partial \ql_+} \left\| \qs^{rk}_0-\qs^{rk}_1 \right\|_1=0$ 
and \\
$\sqrt{\ql_-} w_1 w_2 \frac{\partial}{\partial \ql_-}  \left\| \qs^{rk}_0-\qs^{rk}_1 \right\|_1=0$
can both easily be written in the form\\
 $\frac{w_2}{w_1}=expression$.
Equating these two expressions gives us another hyperbola,
\be
	\left( d^2\ql_+ + \frac{d^2}{2} \ql_- -d \ql_- -2(1-\qb) \right) \left( d^2\ql_- 
	+ \frac{d^2}{2} \ql_+ -d\ql_+ -2\qb \right)
	-\ql_-\ql_+ (d-\frac{d^2}{2})=0.
\ee
The stationary point lies at the crossing of these two hyperbolas. There are four crossing points,
\bea
\ql_+=0 &;\quad\quad & \ql_-=\frac{4(1-\qb)}{d(d-2)}\\
\ql_+=\frac{4\qb}{d(d-2)} &;& \ql_-=0\\
\label{eq:sol3}
\ql_+=\frac{1}{2d(d-1)}+\frac{1-2\qb}{d^2}&;&\ql_-=\frac{1}{2d(d-1)}-\frac{1-2\qb}{d^2}\\
\label{eq:sol4}
\ql_+=\frac{2+d(1-2\qb)}{2d^2} &;& \ql_-=\frac{2-d(1-2\qb)}{2d^2}.
\eea
In the steps above, we have multiplied our derivatives
by $\ql_+$, $\ql_-$, $w_1$ and $w_2$; this has introduced spurious zeros that now need to be removed.
From (\ref{eq:ddx},\ref{eq:ddy}) it is easily seen that $\ql_+=0$ and $\ql_-=0$ are never stationary points 
since the derivatives diverge near these values.  
Furthermore, we find that substitution of (\ref{eq:sol4}) into the derivatives does not yield two zeros.
Expression (\ref{eq:sol3}) is the only stationary point.
As the function value lies higher there than in other points, we conclude that $\left\| \qs^{rk}_0-\qs^{rk}_1 \right\|_1$
is concave.

{\bf Constrained optimization}.
The optimization problem is constrained by the fact that the $\ql$ eigenvalues are non-negative. 
For $\qb \geq \qb_{\rm sat}$ the stationary point satisfies the constraints and hence is the 
optimal choice for $\qb \geq \qb_{\rm sat}$.

For $\qb<\qb_{\rm sat}$ the stationary point has 
$\ql_- <0$, i.e.\,it lies outside the allowed region.
Because of the concavity 
the highest function value which satisfies the constraints occurs 
at $\ql_0=0$, $\ql_1=0$, $\ql_+=0$ or $\ql_-=0$. 
It is easily seen that
$\ql_0 \geq 0$ implies $\ql_+ \leq \frac{1}{d-1}-\frac{2\qb}{d}$ and $\ql_1 \geq 0$ implies $\ql_+ \leq \frac{4\qb}{d(d-2)} - \frac{d}{d-2}\ql_-$ and $\ql_- \leq \frac{4\qb}{d^2} -\frac{d-2}{d}\ql_+$. 
In the range $\qb < \qb_{\rm sat}$ it holds that $\frac{4\qb}{d(d-2)}<\frac{1}{d-1}-\frac{2\qb}{d}$; 
hence the $\ql_0$-constraint is irrelevant in this region. We get $\ql_1=0$ when $\ql_+=\frac{4\qb}{d(d-2)} - \frac{d}{d-2} \ql_-$. 
Substitution gives 
$\frac{1}{2} \left|\left| \qs^{rk}_0-\qs^{rk}_1 \right|\right|_1 =\\
\frac{\sqrt{2}}{d-2} \sqrt{2(1-\qb)+d (1-2\beta +d\left( 1-2\qb (d-1)\ql_-) \right) \left(d^2 \ql_--4\beta \right)}$ which has its maximum at $\ql_-=0$ for non-negative values of $\ql_-$. So either $\ql_-=0$ or $\ql_+=0$.
This leaves two options for the maximum at low~$\qb$,
\bea
\ql_+=0\quad;\quad \ql_- = \frac{4\qb}{d^2} \quad &\Rightarrow&\quad \frac{1}{2} \left\| \qs^{rk}_0-\qs^{rk}_1 \right\|_1 =  0.\\
\ql_- = 0\quad;\quad \ql_+=\frac{4\qb}{d(d-2)} \quad &\Rightarrow& \quad \frac{1}{2} \left\| \qs^{rk}_0-\qs^{rk}_1 \right\|_1 =
 2 \sqrt{2} \frac{\sqrt{\qb(d-2)-2\qb^2(d-1)}}{d-2}.
\label{eq:opt2}
\eea
Clearly (\ref{eq:opt2}) is the larger of the two and therefore the optimal choice.
\hfill$\square$


\bibliographystyle{unsrt}
{\small
\bibliography{poging3}
}

\end{document}